\documentstyle[11pt]{article}
\input epsf
\def\thefootnote{\fnsymbol{footnote}}

\newcommand{\rll}{\rule[-0.3cm]{0cm}{0.9cm}}

\newcommand{\nn}{\nonumber}

\newcommand{\ri}{\right}
\newcommand{\lf}{\left}

\newcommand{\ep}{\varepsilon}
\newcommand{\eq}{\begin{equation}}
\newcommand{\en}{\end{equation}}
\newcommand{\bea}{\begin{eqnarray}}
\newcommand{\eea}{\end{eqnarray}}

\newcommand{\ba}{\begin{array}}
\newcommand{\ea}{\end{array}}

\newcommand{\CC}{{\hbox{\rm C\kern-0.5em{$\sf I$}}}}
\newcommand{\II}{\hbox{{\rm l{\hbox to 1.5pt{\hss\rm l}}}}}
\newcommand{\RR}{{\hbox{$\rm\textstyle I\kern-0.2em R$}}}
\newcommand{\ZZ}{{\hbox{$\sf\textstyle Z\kern-0.4em Z$}}}

\newcommand{\CM}{{\cal{M}}}

\newcommand{\resection}[1]{\setcounter{equation}{0}\section{#1}}

\newcommand{\half}{\frac{1}{2}}

\newcommand{\D}{{\cal D}}
\newcommand{\E}{{\cal E}}

\newcommand{\R}{{\cal R}}

\newcommand{\C}{{\cal C}}

\newcommand{\NP}[1]{Nucl.\ Phys.\ {\bf #1}}
\newcommand{\PL}[1]{Phys.\ Lett.\ {\bf #1}}

\newcommand{\CMP}[1]{Comm.\ Math.\ Phys.\ {\bf #1}}
\newcommand{\CPC}[1]{Computer Phys.\ Comm.\ {\bf #1}}
\newcommand{\PR}[1]{Phys.\ Rev.\ {\bf #1}}

\newcommand{\MPL}[1]{Mod.\ Phys.\ Lett.\ {\bf #1}}
\newcommand{\IJMP}[1]{Int.\ J.\ Mod.\ Phys.\ {\bf #1}}

\newcommand{\AlBZ}{Al.B.Zamolodchikov}

\newcommand{\JSP}[1]{J.\ Stat.\ Phys.\ {\bf #1}}
\newcommand{\JP}[1]{J.\ Phys.\ {\bf #1}}

\newcommand{\ubl}[1]{\{#1\}}
\newcommand{\usbl}[1]{\left(#1\right)}
\newcommand{\iint}{\int^{\infty}_{-\infty}}
\newcommand{\iintd}{\int^{\infty}_{-\infty}\! d\theta}

\newcommand{\wtilde}{\widetilde}

\newcommand{\widetable}{\renewcommand{\arraystretch}{1.65}}

\newcommand{\fract}[2]{{\textstyle\frac{#1}{#2}}}

\newcommand{\halft}{\fract{1}{2}}
\newcommand{\prtial}{\frac{\partial}{\partial\theta}}

\newcommand{\logc}{{\rm Log}_{\C}}
\newcommand{\logca}{{\rm Log}_{\C_a}}
\newcommand{\logea}{{\rm Log}_{\E_a}}
\newcommand{\logci}{{\rm Log}_{\C_1}}
\newcommand{\logcii}{{\rm Log}_{\C_2}}

\newcommand{\opnup}[1]{\renewcommand{\\}{\\[50 pt]}}
\renewcommand{\bar}{\overline}
\renewcommand{\tilde}{\widetilde}
\newcommand\rdilog{{\cal L}}

\newcommand\re{\hbox{Re}\,}
\newcommand\im{\hbox{Im}\,}
\newcommand\sign{\hbox{Sign}\,}

\renewcommand\hat{\widehat}
\newcommand\epsAnum{4}
\begin{document}
\begin{titlepage}
\vskip 0.5cm
\begin{flushright}
DTP-97/29 \\
June 1997 \\
hep-th/9706140
\end{flushright}
\vskip 1.5cm
\begin{center}
{\Large {\bf Excited states in some simple perturbed}} \\[5pt]
{\Large {\bf conformal field theories } }
\end{center}
\vskip 0.9cm
\centerline{Patrick Dorey and Roberto Tateo}
\vskip 0.6cm
\centerline{\sl Department of Mathematical Sciences,  
}
\centerline{\sl  University of Durham, Durham DH1 3LE, 
England\,\footnote{e-mail: {\tt P.E.Dorey@durham.ac.uk,
Roberto.Tateo@durham.ac.uk}} }
\vskip 1cm
\begin{abstract}
\vskip0.2cm
\noindent
The method of analytic continuation is used to find exact integral
equations for a selection of finite-volume energy levels for the
non-unitary minimal models ${\cal M}_{2,2N+3}$ perturbed by their
$\varphi_{13}$ operators. The $N{=}2$ case is studied in particular
detail. Along the way, we find a number of general results
which should be relevant to the study of excited states in other
models.

\bigskip

\noindent PACS numbers: 05.50+q, 11.25.Hf, 64.60.Ak, 75.10.Hk
\end{abstract}
\end{titlepage}

\setcounter{footnote}{0}
\def\thefootnote{\arabic{footnote}}

\resection{Introduction}
The finite-volume spectrum of a quantum field theory encodes a 
great deal of information, 
of interest even if the ultimate objective is to understand
the behaviour of the theory in a
world of infinite extent~\cite{La,Lb}. In the unusually tractable
examples provided by
integrable theories in 1+1 dimensions, it has been
known for some time that an effective way to obtain the lowest level of
this
spectrum (the ground-state energy) is provided by the integral
equations of the thermodynamic Bethe ansatz, or TBA~\cite{Zb}. More
recently, modifications to these equations have been found which
extend the treatment to other (`excited') energy
levels, for the rather simple but nonetheless nontrivial case of the
scaling Lee-Yang model, or SLYM~\cite{BLZa,DTa}. 
(An alternative approach, based on the
Destri -- de Vega equation, was initiated in refs.~\cite{FMQRa},
but its relation with the TBA remains to be understood.)
In this paper we explore how the methods of~\cite{DTa} can be
generalised to a series of models,
namely the perturbations of the minimal models
$\CM_{2,2N+3}$ by their $\varphi_{13}$ operators. Apart from $N{=}1$,
which is just the scaling Lee-Yang model, these theories all have more
than one mass in their spectrum, and their excited-state TBA
equations exhibit a number of new phenomena. 
The bulk of the paper concerns the $N{=}2$ case, for which we have some
detailed numerical results. For higher values of $N$, our results are
less complete, though we can be confident that we have captured
the exact behaviour of the one-particle levels in the infrared region.

The paper begins with
a brief review of the relevant perturbed conformal
field theories, and a discussion of some general properties of TBA
equations that will be important later. Ref.~\cite{DTa} can be
consulted for some further background on the idea of analytic continuation as
applied to TBA systems. In section~3 the method is used to give a
reasonably complete analysis of the first few levels for $N{=}2$, whilst
section~4 describes some features which emerge when the same
techniques are applied to the other theories. None of these models is of
overwhelming physical importance,
and our main aim has been to use them as a
testing-ground for the whole analytic continuation approach to generalised 
TBA equations. With this in mind, the concluding section~5 includes a
summary of those features which have emerged during our investigations
that should have wider applicability. There are two appendices:
the first recalls some field-theoretic predictions for infrared
asymptotics, and the second compares the numerical solutions of our
equations with results obtained from the truncated conformal space
approach.

\resection{The $T_N$-related perturbed conformal field theories}
The minimal models $\CM_{2,2N+3}$ have $N{+}1$ primary 
fields $\varphi_{1,s}\,$, 
occupying a single row of the 
Kac table. The central charge is
$c=-2N(6N{+}5)/(2N{+}3)$, and the 
scaling dimensions of the primary fields are
\[
d_{1,s}=-\frac{(s{-}1)(2N{+}2{-}s)}{2N{+}3}
\qquad s=1\dots N{+}1~.
\]
Perturbing by $\varphi_{13}$ results in a factorised scattering
theory sometimes denoted by $T_N$, or else $A^{(2)}_{2N}$~\cite{CMa,Sa}.
This has $N$ self-conjugate particles with masses $M_a$:
\[
M_a=\frac{\sin(\pi a/h)}{\sin(\pi/h)}M_1
\qquad ~~~a=1\dots N~,
\]
where $h=2N{+}1$, and two-particle S-matrix elements $S_{ab}(\theta)$:
\eq
S_{ab}=
 \prod^{ \atop a+b-1}_{|a-b|+1\atop {\rm
step~2}}\ubl{l}\ubl{h-l}   
\qquad (a,b=1\dots N)~;
\label{sdef}
\en
\[
\ubl{x}=\usbl{x-1}\usbl{x+1}\quad;\quad
\usbl{x}(\theta)={\sinh\bigl({\theta\over 2}+{\imath\pi x\over 2h}\bigr)\over
        \sinh\bigl({\theta\over 2}-{\imath\pi x\over 2h}\bigr)}~.
\]
The exact relation between the $\varphi_{13}$ coupling $\lambda$
and the mass $M_1$ of the lightest particle was found by
\AlBZ~\cite{Zg}. Setting
$g(x)=\sqrt{\Gamma(x)/\Gamma(1{-}x)}\,$, it is
\eq
M_1(\lambda)=\frac{4\sin\frac{\pi}{h}}{\sqrt{\pi}\,}
\frac{\Gamma\!\lf(\frac{1}{h}\ri)}{\Gamma\!\lf(\frac{1}{2}{+}\frac{1}{h}\ri)}
\lf[\pi\frac{(h{-}2)(h{-}4)}{\,(h{+}2)^2}
g\!\lf(\fract{h-4}{h+2}\ri)g\!\lf(\fract{h}{h+2}\ri)
\ri]^{\frac{h+2}{4h}}\!\!\lf(-\lambda\ri)^{\frac{h+2}{4h}}\,,
\label{mrel}
\en
In the normalisations implicit in \cite{Zg},
a sensible scattering theory with real masses for the particles 
is obtained when the coupling $\lambda$ is real and {\it negative}. The 
only exception is $N{=}1$, where the imaginary values of some conformal 
structure constants mean that a real perturbative expansion is obtained if
$\lambda$ is purely imaginary, and $\im\lambda$ must be positive for real
masses in the infrared~\cite{Zb,YZa}.

Since the S-matrix (\ref{sdef}) is diagonal, the TBA equations 
for the ground state energy on a circle of circumference $R$
are~\cite{Zb,KMa}:
\eq
\ep_a(\theta)=m_ar\cosh\theta-\sum_{b=1}^N\phi_{ab}{*}L_b(\theta)~,
\label{zeroTBA}
\en
with
\[
r=M_1R\,,~\,
m_a={M_a\over M_1}\,,~\,
\phi_{ab}(\theta)=-\imath\prtial\log S_{ab}(\theta),~\,
L_a(\theta)=\log\lf(1{+}e^{-\ep_a(\theta)}\ri),
\]
and $f{*}g(\theta)$ denoting the convolution
$\frac{1}{2\pi}\iintd' f(\theta{-}\theta')g(\theta')\,$.
Solving these equations yields $N$ pseudoenergies $\ep_a(\theta)\,$; 
the ground state energy is then
\[
E_0(\lambda,R)=E_{\rm bulk}(\lambda,R)-\frac{\pi}{6R}c(r)~;
\]
\eq
E_{\rm bulk}(\lambda,R)=-\frac{M_1(\lambda)^2}{8 \sin(2 \pi /h)}R~,\quad
c(r)=\frac{3}{\pi^2}\sum_{a=1}^{N}\iintd\,r\cosh\theta L_a(\theta)~.
\label{zeroTBAc}
\en
It will sometimes be convenient to work with the scaling functions
$F_i(r)$, related to the energy levels $E_i$ by
$E_i(\lambda,R)=\frac{2\pi}{R} F_i(M_1R)$.

The functions $Y_a(\theta)=\exp(\ep_a(\theta))$ 
provide an $r$-dependent set of
solutions to a set of functional relations called a 
Y-system~\cite{Zf,RTVa}
\eq
Y_a(\theta-\fract{\imath\pi}{h})Y_a(\theta+\fract{\imath\pi}{h})=
\prod_{b=1}^N\lf(1+Y_b(\theta)\ri)^{l^{[T_N]}_{ab}},
\label{Ysys}
\en
where $l^{[T_N]}_{ab}=\delta_{a,b{-}1}+
\delta_{a,b{+}1}+\delta_{a,N}\delta_{b,N}$ is the incidence matrix of
the $T_N$ graph. A (none-too-obvious) consequence of these equations
is that the functions $Y_a(\theta)$ are $\imath\pi(h{+}2)/h$-periodic:
\eq
Y_a(\theta+\fract{\imath\pi(h{+}2)}{h})=Y_a(\theta)~.
\label{Yper}
\en
Subtracting the right-hand side of (\ref{Ysys}) from the left-hand
side yields an analytic function of $r$ and $\theta$ which is
identically zero.
This function therefore remains zero during any process of analytic
continuation,
and so the Y-system and all of its consequences hold equally for all the
continuations of the basic TBA to be encountered below. With
this in mind, it is worth pausing to record a couple of other
general properties before proceeding to specific examples.

The first concerns the behaviour of the
$Y_a$ as functions of $r$ and $\theta$ combined. If we define a
new pair of variables as follows:
\eq
a_+=\left(re^{\theta}\right)^{2h/(h{+}2)}\quad;\quad
a_-=\left(re^{-\theta}\right)^{2h/(h{+}2)}
\label{aa}
\en
then we claim that each $Y_a(a_+,a_-)$ can be expanded as a power series
about $a_+=a_-=0$, with a finite 
domain $\D\subset\CC\times\CC$\ of convergence. This 
is a conjecture, but note 
that it corresponds to an expansion about the central region in the
`kink limit' of the TBA, where we would not expect to find any
singularities, all pseudoenergies tending to constants there.

The second general property follows from the first. Consider
simultaneous changes of $r$ and $\theta$ which leave $a_+$ and $a_-$
invariant. These take the form $r\rightarrow e^{\imath\alpha}r$;
$\theta\rightarrow\theta{+}\imath\beta$ with $\alpha=(p{+}q)\pi(h{+}2)/2h$,
$\beta=(p{-}q)\pi(h{+}2)/2h$, and $(p,q)$ a pair of integers. The $(-1,0)$
case of this transformation shows that
\eq
Y_a(e^{-\imath\pi(h{+}2)/2h}r,\theta{-}\imath\pi\fract{h{+}2}{2h})
=Y_a(r,\theta)\quad (\,(a_+,a_-)\in\D\,)\,.
\label{ytrans}
\en
This result is only guaranteed while $|r|$ and $|\theta|$ are small enough
that $(a_+,a_-)$ lies in $\D$. However as functions of $\theta$ alone,
the $Y_a$ are expected to be entire (in the conformal case
this can be proved in some situations~\cite{BLZa}\,).
If $|r|$ is sufficiently small, then
$(a_+,a_-)$ 
will lie in $\D$ for, say, real $\theta$ in the unit interval; the identity
(\ref{ytrans}) holds for these values of $\theta$, and can then be 
analytically continued to all $\theta$ while $r$ remains fixed. 
Using the real analyticity of the $Y$'s we can then deduce that for 
these sufficiently small values 
of $|r|$ the functions $Y_a(r,\theta)$ have the symmetry 
\eq
Y_a(r,\theta)=Y_a^*(\tilde r,\tilde\theta)
\label{selfconj}
\en
where
\eq
\tilde r = e^{\imath\pi(h{+}2)/2h}r^*\quad,\quad
\tilde\theta=\imath\pi\fract{h{+}2}{2h}+\theta^*
\label{tilddef}
\en
and the asterisks denote complex conjugation.
For larger values of $|r|$, the point $(a_+,a_-)$ will be outside $\D$ for all
$\theta$, and the argument breaks down. Indeed,
the functions $Y_a(r,\theta)$ acquire a multi-valued structure reflecting
the multivalued nature of the energy levels themselves. 
However, since all of these functions
can be found by analytic continuation from the small-$|r|$
region, a memory of the earlier relation remains in an involution on the
set of all $Y$'s: given a function $Y_a(r,\theta)\,$, the exponential of
the solution to some
(perhaps generalised) TBA equation, another function $\tilde Y_a(r,\theta)$
can be defined by
\eq
\tilde Y_a(\tilde r,\theta)=Y^*_a(r,\tilde\theta)~.
\label{yconj}
\en
The logarithm of this function, $\tilde\ep_a(\theta)=\log\tilde Y_a(\tilde
r,\theta)\,$, will again solve a (possibly different)
TBA equation, with the radius $r$ replaced by the reflected radius
$\tilde r$. 
The observation turns out to be rather useful, allowing solutions of
the generalised TBA equations to be obtained in some situations where
a direct numerical attack fails. 
So as not to have to rely for it
on the power-series expansion, we also checked its validity, for
$N{=}2$, $3$ and $4$,
in various situations where the equations related to both $Y_a$ and $\tilde
Y_a$ could be solved numerically.

\begin{figure}[htbp]
\epsfxsize=11.5cm
\epsfbox{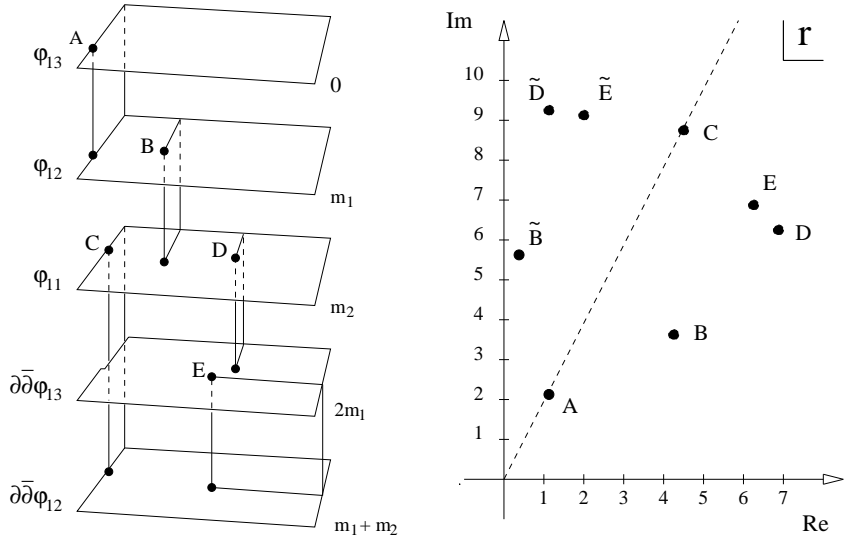}
\caption{ 
\leftskip=.5cm  {\protect \small  
 The first five sheets of the Riemann surface 
of the scaling function $F(r)$, showing on the left the connectivity of
the various sheets, and on the right the positions of the square-root
branch points over the complex $r$-plane.}
\rightskip=0.5cm }  
\label{surface}
\end{figure}

\resection{Excited states for $\CM_{2,7}+\lambda\varphi_{13}$}
In this section we restrict our attention to the case $N{=}2$, $h{=}5$.
The approach proposed
in~\cite{DTa} requires the continuation of $r$ along some path in the
complex plane, in the hope of inducing a monodromy in $c(r)$ which will
convert the ground state into an excited state. To decide which paths to
take, we used the truncated conformal space approach (TCSA)~\cite{YZa},
and in particular the program of ref.~\cite{LMa},
to map the first few sheets of the Riemann surface associated with the ground
state scaling function $F_0(r)$. The results are shown schematically in
figure~\ref{surface}.

\setlength{\unitlength}{1.mm}

\begin{figure}[htbp]
\epsfxsize=11.5cm
\epsfysize=7.0cm
\epsfbox{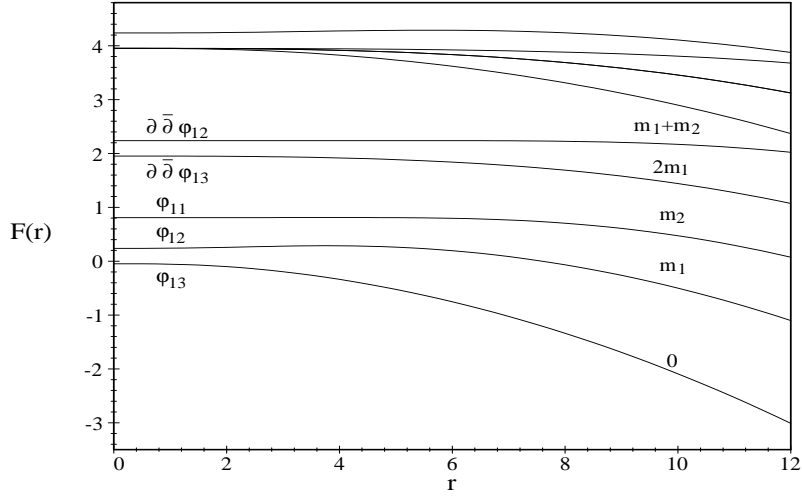}
\caption{  \leftskip=.5cm {\protect \small  
TCSA data for real $r$ between $0$ and $12$, with the lines
labelled both by the conformal fields found in the ultraviolet,
and by the mass gaps exhibited as $r\rightarrow\infty$.}
\rightskip=0.5cm }  
\label{TCSA2}
\end{figure}
\begin{figure}[htbp]
\epsfxsize=11.5cm
\epsfysize=7.cm
\epsfbox{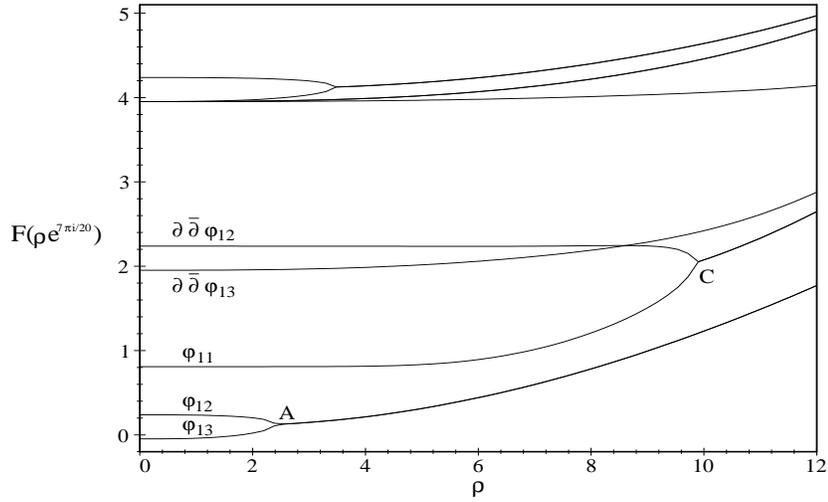}
\caption{ \leftskip=.5cm  {\protect \small  
TCSA data for complex $r$ on the positive-$\lambda$ line $r=\rho
e^{7\pi\imath/20}$, $0\le\rho\le 12$. Three branch points can be seen;
those marked $A$ and $C$ also appear on figure~1.}
\rightskip=0.5cm }  
\label{TCSA3}
\end{figure}

Figures 2 and 3 may also help to visualise how the sheets fit together.
Figure~2 plots various $F(r)$ for real values of $r$ between $0$ and
$12$, labelling the lines both by their ultraviolet limits near $r{=}0$,
and by their asymptotic mass gaps at large $r$. Figure~3 shows the real
parts of these same
$F(r)$ along another line of interest, namely $r=\rho e^{7\pi\imath/20}$, 
$\rho\in\RR^+$. Via equation (\ref{mrel}) and the relation
$r=M_1(\lambda)R$, this line also corresponds to real $R$ and $\lambda$,
but now with $\lambda$ positive. Consideration of the power series
expansions in $\lambda$
provided by perturbed conformal field theory
shows that the scaling functions
must be real here too on some initial
segment of the positive-$\lambda$
line. Beyond the radius of convergence of the perturbative expansions
singularities are found, after which the
$F(r)$ can pick up imaginary parts. Examples of singularities on the
positive-$\lambda$ line are the
branch points marked $A$ and $C$ in figures 1 and~3.

As noted in~\cite{DTa}, another consequence of the perturbative 
series is that all other branch points
occur in pairs $r$, $\wtilde r$, related by a complex conjugation 
of $\lambda$, which reflects the complex $r$-plane in the positive-$\lambda$
line. This parallels the involution of the $Y$'s discussed
at the end of the last section. For $N{=}2$, the relationship is
\eq
\wtilde r= e^{7\pi\imath/10}r^*~.
\label{rrel}
\en
Examples in figure~1 are the pairs $B,\wtilde B$; $D,\wtilde D$; and
$E,\wtilde E$. 

Using the branch points, it is possible to
move between the energy levels. 
A path encircling $A$ will link the ground state with the first
excited state, one encircling $A$ and then $B$ will link the ground
state with the second excited state, and so on. During this
continuation, singularities in one or other of 
$L_1$ and $L_2$ may cross the real axis. 
Whenever this happens, all integration contours in the TBA equations
must be distorted
to run around these singularities. When the contours are returned
to the real axis, residue terms are picked up, resulting in 
modified equations in which the positions of some singularities
make an explicit appearance. We will call these singularities 
`active', and the others `inactive'.

This was described for the case of the SLYM in~\cite{DTa}.
The principles are the same here, but the practice is
considerably more complicated, and the rest of this section will
be devoted to a detailed description of how it goes for the cases
of zero, one and two pairs of active singularities (note, the
$\theta\rightarrow-\theta$ symmetry of the initial TBA means
that singularities always either appear or disappear from the
equations in pairs).
It will sometimes be convenient to
use the functions $z_a(\theta)$,
$a=1,2$, where
\[
z_a(\theta)=1+e^{-\ep_a(\theta)}=1+Y_a(\theta)^{-1}\,.
\]
Singularities of $L_a(\theta)=\log z_a(\theta)$ are found at zeroes and 
poles of $z_a$, where $Y_a$ takes the values $-1$ and $0$
respectively. It will be important later that these singularities
never come singly: the $T_2$ Y-system (\ref{Ysys}) ties together the
values of the $Y_a(\theta)$ at locations separated by integer multiples
of $\imath\pi/5$. 
We will label the zeroes of $Y_1(\theta)$ as $\pm\beta_2^{(j)}$,
$j=0\dots\infty$. With each zero, the following other special values are
forced:
\vskip -4pt
\eq
\hbox{
\renewcommand{\arraystretch}{1.25}
\begin{tabular}{|c|ccccc|}
\hline
$k$ &{}~~$\dots$&{}~~$-1$~~ & 
{}~~$0$~~ & {}~~$1$~~ & {}$\dots$~~{}  \\
\hline
$Y_1(\beta^{(j)}_2+{\imath k\pi\over 5})$& 
 {}~~ $\dots$ & $y$  &$0$ & $1/y$ & $\dots$~~{}\\
$Y_2(\beta^{(j)}_2+{\imath k\pi\over 5})$&
 {}~~ $\dots$ & $-1$ &$0$ & $-1$  & $\dots$~~{}\\
\noalign{\vskip 1pt}
\hline
\end{tabular}
}
\label{y1tab}
\en
\noindent
Here $y$ is free, although it should not be zero since, as mentioned
earlier, $Y_1$
and $Y_2$ are expected to be entire functions of $\theta$.
The zeroes of $Y_2(\theta)$ occur either at the points $\pm\beta^{(j)}_2$
already captured in
the above table, or else at a further set of points $\pm\beta^{(j)}_1$,
which entail a different singularity pattern:
\vskip -4pt
\eq
\hbox{
\renewcommand{\arraystretch}{1.25}
\begin{tabular}{|c|ccccc|}
\hline
$k$ &{}~~$\dots$&{}~~$-1$~~ & 
{}~~$0$~~ & {}~~$1$~~ & {}$\dots$~~{}\\
\hline
$Y_1(\beta^{(j)}_1+{\imath k\pi\over 5})$ & {}~~$\dots$ 
& $-1$  &$a$ & $-1$ & $\dots$~~{}\\
$Y_2(\beta^{(j)}_1+{\imath k\pi\over 5})$&
 {}~~$\dots$ & $y$ &$0$ & $(1{+}a)/y$  &$\dots$~~{}\\
\noalign{\vskip 1pt}
\hline
\end{tabular}
\label{y2tab}
}
\en
\noindent
This time both $y$ and $a$ are nonzero (the latter condition preventing
singularities in some of the omitted terms in the table).

As for the other type of singularity in $L_a(\theta)$, corresponding
to a zero of $z_a(\theta)$ with $Y_a(\theta)=-1$, 
it is not hard to check that $z_2(\theta)=0$ 
forces (\ref{y1tab}), with $\beta^{(j)}_2$ equal to either
$\theta{-}\imath\pi/5$ or $\theta{+}\imath\pi/5$,
while $z_1(\theta)=0$ leads to (\ref{y2tab}),
this time with $\beta^{(j)}_1$ equal to either
$\theta{-}\imath\pi/5$ or $\theta{+}\imath\pi/5$. This is the reason for
the at-first-sight perverse labelling: 
for the analytic continuation of the
TBA equations, it is
the zeroes of the $z_a(\theta)$, located at
$\beta^{(j)}_a\pm\imath\pi/5$, that turn out to
play the crucial role.

To unravel the singularity structure of any particular 
solution,
it often helps to reconstruct its `history' through the analytic
continuation, remembering the $\theta\rightarrow-\theta$
symmetry of the equations. Also useful are the facts that whenever $r$ is
real, the singularity pattern must additionally
be symmetrical under the standard conjugation
$\theta\rightarrow\theta^*$, and that whenever $r$ lies on
the positive-$\lambda$ line, so that $\tilde r=r$ in (\ref{tilddef}),
the pattern is either symmetrical under the shifted conjugation
$\theta\rightarrow\tilde\theta$,
or else is mapped by this operation onto the pattern 
for another solution $\tilde Y_a$.

\begin{figure}[tbh]
\hskip 1.3cm
\epsfxsize=10.5cm
\epsfbox{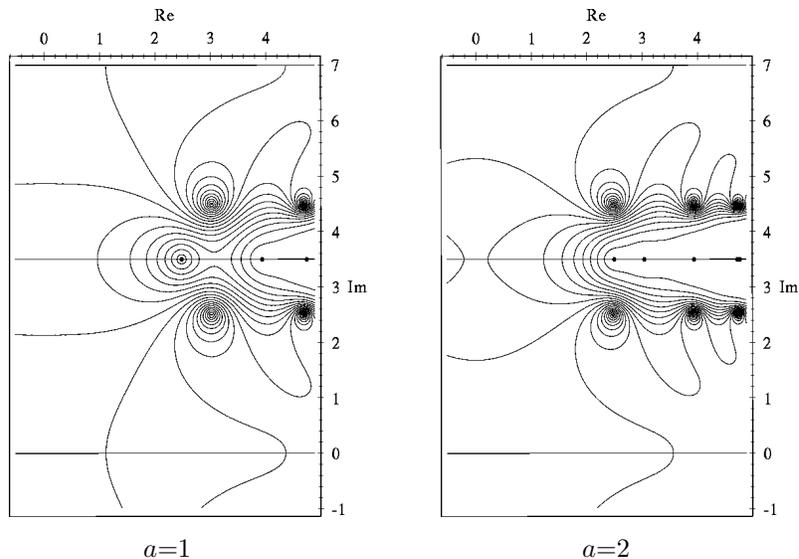}
\vskip 1pt
{\small\noindent\hfil$a{=}1$\qquad\qquad\hfil\qquad$a{=}2$\hfil\break}
\vskip -10pt
\caption{  \leftskip=.5cm {\protect \small  
Contour plots of $|z_a(\theta)|/(1{+}|z_a(\theta)|)$ in the complex
$\theta$ plane, for the basic TBA equation with $r{=}1$. Concentric
patterns of
contours are signs of singularities in $L_a(\theta)$. Those 
corresponding to zeroes of $Y_a(\theta)$
are marked by dots $\bullet\,$; the others are zeroes of $z_a(\theta)$.
The scales on the axes are in units of $\pi/5$.}
\rightskip=0.5cm }  
\label{epsA}
\end{figure}

For later use, we will record the situation for the solutions
to the basic TBA equation (\ref{zeroTBA}), when $r$ is real. 
A typical pattern, in fact that
found at $r{=}1$, is shown in figure~\ref{epsA}.
The points $\pm\beta_a^{(j)}$ all have
imaginary parts equal to $7\pi/10$ 
(the same, by the periodicity (\ref{Yper}), 
as $-7\pi/10\,$), and so
all of the zeroes of $Y_1$ and $Y_2$ lie on the line $\im\theta
=\pm 7\pi/10$. The zeroes of $z_1$ and $z_2$, as follows from
(\ref{y2tab}) and (\ref{y1tab}) respectively, lie on the lines $\im\theta
=\pm\pi/2$.
For definiteness, we will choose the $\beta$'s to have negative
real parts, and to be ordered from left to right:
\eq
\dots<\re\beta^{(j)}_a<\re\beta^{(j-1)}_a<\dots<\re\beta^{(0)}_a<0~.
\label{betapat}
\en
As $r\rightarrow 0$, solutions to TBA equations of
this sort split into pairs of `kink systems'.
One starts near $-\log(1/r)$ and runs
leftwards, the other starts near $+\log(1/r)$ and runs
rightwards, and in the region between the two the pseudoenergies are
approximately constant. With the labelling choice just adopted,
the zeroes of $Y_1$ and $Y_2$ in the left-hand system are
at $\{\beta_2^{(i)}\}$ and $\{\beta^{(i)}_2,\beta_1^{(j)}\}$ 
respectively, and those of the right-hand system are at
$\{-\beta_2^{(i)}\}$ and $\{-\beta^{(i)}_2,-\beta_1^{(j)}\}$. Thus the
singularities visible in figure~\ref{epsA} all belong to the developing
right-hand kink system, and move further to the right as $r$
decreases.

In the deep ultraviolet,
the mutual interaction of the two kink systems becomes vanishingly small, 
and they separately solve kink forms of the TBA equations, obtained by 
replacing every occurrence of $r\cosh\theta$ either by $\half re^{-\theta}$
(for the left-hand kink system) or by $\half re^{\theta}$ (for the
right-hand system). In the kink forms of the equations, the dependence 
on $r$ is trivial: any further changes can be absorbed by a compensating 
shift in $\theta$. This tells us something new about the singularity
locations in this limit: the singularities in
the left-hand set are at fixed distances from 
the point $\theta=-\log(1/r)$, while those in
the right-hand set are oppositely placed, at fixed distances from the 
point $\theta=\log(1/r)$. To make this more precise, we have
\eq
\lim_{r\rightarrow 0}
\lf[\beta^{(j)}_a+\log(1/r)\ri]=c^{(j)}_a~,
\label{cform}
\en
with $\{c^{(j)}_a\}$ a set of constants satisfying $\im
c^{(j)}_a=\pm7\pi/10$. 
At large $j$, the real parts of the $c^{(j)}_a$ are given by 
\[
\re c^{(j)}_a\sim-\log\lf(\fract{2}{m_a}(2j+1)\pi\ri)\qquad (j\gg 1)~.
\]
For smaller $j$ these real parts are somewhat distorted,
though the counting (starting at $j=0$) is correctly given by this
formula.

During analytic continuation, the singularities will move around,
and for most of the time
we will use the same labels for the singularities of
the continued solutions. To avoid ambiguity
we must remember the
particular path in the complex $r$-plane
chosen to reach a given solution, since there are 
closed paths on the Riemann surface around which
pairs of singularities swap over. Some examples
will be seen later.

\subsection{The basic TBA in the complex plane}
\label{basicsec}
\begin{figure}[tbh]
\hskip 1.3cm
\epsfxsize=10.5cm
\epsfbox{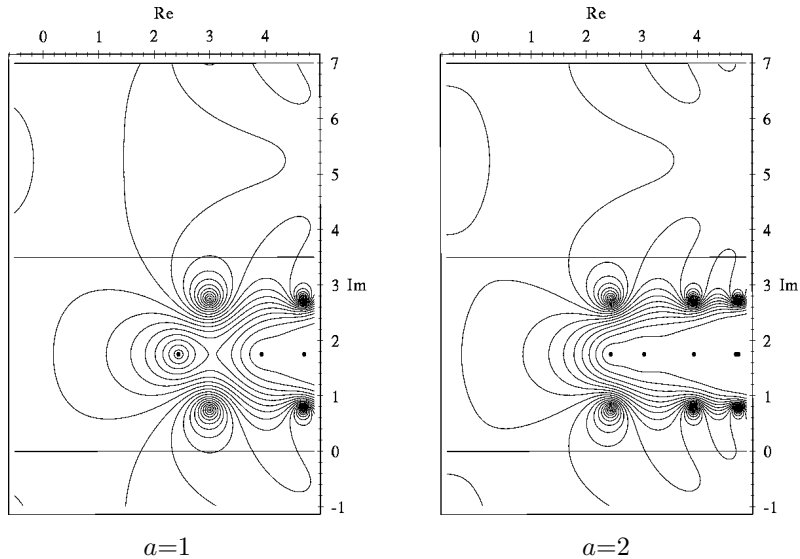}
\vskip 1pt
{\small\noindent\hfil$a{=}1$\qquad\qquad\hfil\qquad$a{=}2$\hfil\break}
\caption{  \leftskip=.5cm {\protect \small  
Contour plots of $|z_a(\theta)|/(1{+}|z_a(\theta)|)$ in the complex
$\theta$ plane, for the ground-state
solution to the basic TBA equation with $r{=}e^{7\pi\imath/20}$,
$\rho{=}1$.
Labelling as in figure~\epsAnum.}
\rightskip=0.5cm }  
\label{epsB}
\end{figure}

This case provides the starting-point for the investigation.
If $r$ is continued up from the real axis onto the positive-$\lambda$
line, then a function $F(\rho e^{7\pi\imath/20})$ is
obtained, from the basic TBA equations 
(\ref{zeroTBA}) and (\ref{zeroTBAc}), which is real out to
$\rho=\rho_0=2.39342(4)$. This point matches the branch point $A$ found from
the TCSA. On this segment
$Y_1$ and $Y_2$ are self-conjugate in the shifted
sense of equations (\ref{tilddef}) and (\ref{yconj}):
\eq
Y_a(\rho e^{7\pi\imath/20},\theta)=
Y_a^*(\rho e^{7\pi\imath/20},\tilde\theta)\,,\quad 0\le\rho\le\rho_0~.
\label{Ysconj}
\en
This means that the singularities in $L_1$ 
and $L_2$ are arranged symmetrically with respect to 
the shifted conjugation
$\theta\rightarrow\tilde\theta=7\pi\imath/10{+}\theta^*\,$.
Here this happens in the
simplest possible way, with the zeroes of $Y_1$ and $Y_2$ lying on
the fixed lines of the mapping, $\im\theta=\pm 7\pi/20$, and the
zeroes of the
$z$'s symmetrically arranged
on the lines $\im\theta=\pm 3\pi/20$ and $\im\theta=\pm 11\pi/20$.
All zeroes in the left half-plane, with negative real parts, lie on the
lower triplet of lines, and all to the right lie on
the upper triplet. 
Figure~\ref{epsB} shows the situation at $|r|=\rho=1$.
Given the symmetry (\ref{Ysconj}), the zeroes
must lie {\it exactly} on the lines just described, if they are not to
be doubled up. A similar phenomenon was observed
in~\cite{DTa} for the SLYM, although at the time it was not understood
in terms of the shifted conjugation symmetry, and so was only verified 
numerically. Such a check was also carried out here,
confirming expectations to a precision of $15$ digits along the whole
segment $0<\rho<\rho_0$, and then showing the singularities moving
away with a square-root
type behaviour as $\rho$ passed $\rho_0$ and the solutions ceased to
be self-conjugate.

In the ultraviolet limit,
$\rho\rightarrow 0$, the splitting into kink systems occurs just as
for real values of $r$. Using the result (\ref{cform}) for the
asymptotic singularity positions, equally valid for these
complex values of
$r$, it is easily seen how the singularities move 
during the continuation from the real axis to the
positive-$\lambda$ line.
For example, the zeroes of $Y_1$ and $Y_2$ in the left-hand system 
move up from the line $\im\theta=-7\pi/10$ to
sit on the line $\im\theta=-7\pi/20$, while those in the
right-hand set
move down from $\im\theta=+7\pi/10$ to $\im\theta
=+7\pi/20$. (If we had instead continued $r$ down into the
lower half-plane, these motions would have been reversed.)
Comparing figures~\ref{epsA} and~\ref{epsB} should clarify the 
relationship between the patterns before
and after continuation up onto the positive-$\lambda$ line.

Of the zeroes of the $z_a(\theta)$, those at $\pm(\beta^{(j)}_a
{+}\imath\pi/5)$ now
lie closest to the real axis, and will be the most important
later. Hence we define
\[
\theta^{(j)}_a=\beta^{(j)}_a+\imath\pi/5~.
\]

This was for the ground-state solution to the basic TBA
equation. But there is another
solution to this equation on the segment of the
positive-$\lambda$ line between $\rho=0$ and $\rho=\rho_0$. This
can be obtained by continuing the solution
already discussed anticlockwise around the point 
$r=\rho_0e^{7\pi\imath/20}$. In the process, no singularities
in $L_1(\theta)$ or $L_2(\theta)$ 
cross the real axis and so no modifications to the TBA equation 
(\ref{zeroTBA}) are required. However the pseudoenergies undergo a 
nontrivial monodromy, and on evaluating the integrals in~(\ref{zeroTBAc}),
a different answer is found which matches perfectly with the first
excited state on the segment $0\le\rho\le\rho_0$, which is the line 
labelled by $\varphi_{12}$ in figure~\ref{TCSA3}.

\begin{figure}[tbh]
\hskip 1.3cm
\epsfxsize=10.5cm
\epsfbox{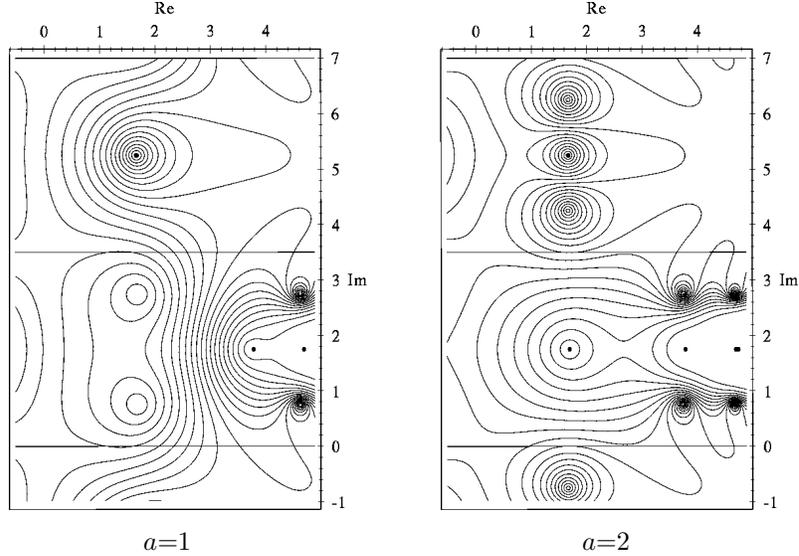}
\vskip 1pt
{\small\noindent\hfil$a{=}1$\qquad\qquad\hfil\qquad$a{=}2$\hfil\break}
\caption{  \leftskip=.5cm {\protect \small  
Contour plots of $|z_a(\theta)|/(1{+}|z_a(\theta)|)$ in the complex
$\theta$ plane, for the excited solution
to the basic TBA equation with $r{=}e^{7\pi\imath/20}$,
$\rho{=}1$.
Labelling as in figure~\epsAnum.}
\rightskip=0.5cm }  
\label{epsD}
\end{figure}

During the anticlockwise continuation of the 
ground-state solution, $\theta^{(0)}_2$, the rightmost zero of 
$z_2(\theta)$ in the left-hand kink system, migrates across to
join the right-hand system, while the matching zero in the right-hand
system makes the opposite journey. 
The final pattern again respects the $\theta\rightarrow\tilde\theta$
symmetry, but now with a different distribution for the imaginary parts
of the singularities between the left and right kink systems.
The change in the pattern of singularities can be appreciated on
comparing figure~\ref{epsD}, which shows the new excited solution on
the positive-$\lambda$ line at $\rho{=}1$, 
with figure~\ref{epsB}, the ground-state solution
at the same value of $\rho$.

\begin{figure}[tbh]
\epsfxsize=12.5cm
\epsfbox{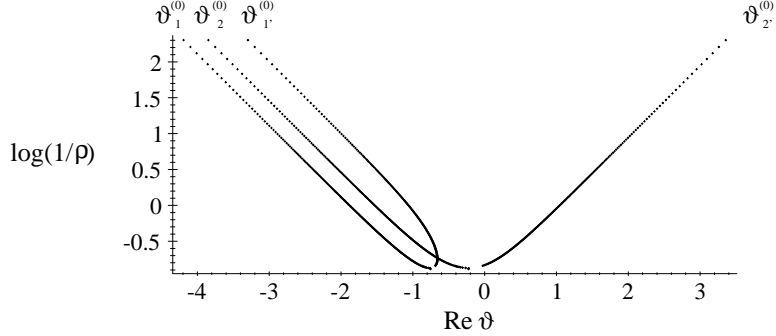}
\caption{\leftskip=.5cm  {\protect \small  
Real parts of $\theta^{(0)}_1$ and $\theta^{(0)}_2$ for $r$
on the positive-$\lambda$ line, plotted against $\log(1/\rho)$. The lines
marked $\theta^{(0)}_1$ and $\theta^{(0)}_2$ are for the ground-state
solution, and those marked $\theta^{(0)}_{1'}$ and $\theta^{(0)}_{2'}$
are for the excited state.}
\rightskip=0.5cm }  
\label{logplot}
\end{figure}

The movement of singularities as the point $\rho=\rho_0$ is approached
along the positive-$\lambda$ line
is shown in figure~\ref{logplot}, which
plots the real parts of $\theta^{(0)}_1$ and $\theta^{(0)}_2$
against $\log(1/\rho)$, both for the
ground state and for the first excited state. The slopes tend
to $\pm 1$ as $\rho\rightarrow 0$, the slopes of $-1$ being
characteristic of members of the left-hand kink system, and the
slope of $+1$ acquired by the `excited' zero of $z_2(\theta)$ showing
that it has joined the right-hand kink system. 
\setlength{\unitlength}{1.mm}

Equation (\ref{cform}) must be modified to take account of the
defection of the two singularities, one from left to right and the
other from right to left.  To prepare for more general situations,
which may involve
further pairs of defectors, we replace
(\ref{cform}) by
\eq
\lim_{|r|\rightarrow 0}\left[\nu^{(j)}_a
\beta^{(j)}_a(r)+\log(1/r)-c^{(j)}_a 
\right]=0~,
\label{fullc}
\en
where $\nu^{(j)}_a$ is $+1$ if the singularity at
$\beta^{(j)}_a$ is still part of the
left-hand kink system, but switches to $-1$ if it has joined
the right-hand system. The values of the constants
$c^{(j)}_a$ are characteristic of
each particular kink solution, while the $\nu^{(j)}_a$ preserve some
memory of how the solution has been obtained by continuation from the
ground state. For example, it appears
that only for the ground state do all of the
$c^{(j)}_a$ have imaginary parts equal to $\pm 7\pi/10$.
For the excited state now under discussion, 
$\im c^{(0)}_2$ has become equal to zero, and
$\nu^{(0)}_2$ is equal to $-1$.

It is instructive to see how the analytic treatment of the ultraviolet
limit must be modified to cope with this excited solution to the basic
TBA. The usual manipulations of the kink forms of
equations (\ref{zeroTBA}) and (\ref{zeroTBAc}) give
\eq
c(0)=\frac{6}{\pi^2}\sum_{a=1}^2 L_+(\C_a)
\label{cdilog}
\en
where, borrowing some notation from of refs.~\cite{KWZa} and \cite{KNa},
\eq
L_+(\C_a)=\frac{1}{2}\int_{\C_a}\left[\frac{\logca(1{+}x)}{x}-
\frac{\logca (x)}{1{+}x}\right]\,dx~.
\label{dilog}
\en
The $\C_a$ are certain contours in the complex plane, and $\logca(z)$
stands for a continuous branch of $\log(z)$ along the contour
$\C_a\,$; the starting-point of the branch will be determined shortly.
The
modification to the ground-state value of $c(0)$ resides in the
form of these contours and branches, and care must be taken to
identify them correctly. The variable change leading to (\ref{cdilog})
tells us that the contour  $\C_a$ in (\ref{dilog}) is given by
$x(\theta){=}\exp(-\ep^{\rm kink}_a(\theta))$, with $\theta$ real and running 
from $+\infty$ (at the start of the contour) to $-\infty$ (at
the end), and $\ep^{\rm kink}_a(\theta)$ the
(excited) kink solution to the TBA equation.
As $\theta\rightarrow\infty$, all pseudoenergies diverge and hence the
contours $\C_a$ always start at the origin. In the opposite direction, 
the kink pseudoenergies tend to finite constants and as a result the
contours terminate at points $x_a=1/Y_a$, with the constants
$(Y_1,Y_2)$ providing a stationary ($\theta$-independent) solution
to the $N{=}2$ case of the Y-system (\ref{Ysys}):
\bea
Y_1^2&=&(1+Y_2)\,,\nn\\
Y_2^2&=&(1+Y_1)(1+Y_2)\,.
\label{Ysystwo} 
\eea
There are four solutions to this equation: one, $(Y_1,Y_2)=(-1,0)$,
appears not to be relevant (it would anyway require $\ep_2=-\infty$),
whilst the others can be written as $(Y_1,Y_2)=(y_1^{(s)},y_2^{(s)})$
with $s=1,3,5\,$, and
\eq
y^{(s)}_a=\frac{\sin(sa\pi/7)\,\sin(s(a{+}2)\pi/7)}{\sin(s\pi/7)^2}~.
\label{Ysystwosoln}
\en
These numbers feature in certain sum rules for dilogarithm functions.
When the dust has settled, it will be most convenient to have these
written in terms of the Rogers dilogarithm $\rdilog(y)$, defined for
real $y$ in the unit interval by
\eq
\rdilog(y)=-\half\int_0^ydt(\frac{\log(1-t)}{t}+\frac{\log t}{1-t})\,.
\en
(In cases when $\C_a$ runs from $0$ to $x$ along the real axis, with all
logarithms in (\ref{dilog}) on their principal branches,
$L_+(\C_a)=\rdilog(x/(1{+}x))\,$.) The sum rules that we shall need
are the following:
\bea
\rdilog\Biggl(\frac{1}{1{+}y^{(1)}_1}\Biggr)
+\rdilog\Biggl(\frac{1}{1{+}y^{(1)}_2}\Biggr)
&=&\phantom{-}\frac{4}{7}\frac{\,\pi^2}{6}
\label{ruleone}\\
-\rdilog\Biggl(1{+}y^{(3)}_1\Biggr)
-\rdilog\Biggl(1{+}y^{(3)}_2\Biggr)&=&-\frac{6}{7}\frac{\,\pi^2}{6}
\label{ruletwo}\\
\rdilog\Biggl(\frac{1}{1{+}y^{(5)}_1}\Biggr)
-\rdilog\Biggl(1{+}y^{(5)}_2\Biggr)&=&\phantom{-}\frac{2}{7}\frac{\,\pi^2}{6}
\label{rulethree}
\eea
Note that all of the arguments have been adjusted so as to lie in the
interval $[0,1]\,$. There do exist prescriptions to extend the Rogers
dilogarithm beyond this range, but at this stage it is better for us
to keep our options open, since the relevant continuations will anyway
be determined by the TBA equations.

For the ground state there is an additional physical requirement that
the $Y$'s be positive and the logarithms real, and this 
selects the $s{=}1$ solution and, via (\ref{ruleone}), the expected
result $c(0)=4/7$ is recovered. For
excited states there are no such constraints, and for the state under
discussion here, $s{=}3$. This can be justified either by an appeal to
numerical results at small $|r|$, or else by noticing that, as a
result of the shifted conjugation symmetry, the functions
$Y^{\rm kink}_a(\theta)=\exp(\ep^{\rm kink}_a(\theta))$ 
are real along the line $\im\theta=21\pi/20$, and that they
each have a single zero on this line, at
$\beta^{(0)}_2$. Since as $\theta\rightarrow\infty$ both are positive
(tending to $+\infty$), they must both be negative at
$\theta=-\infty$, and this picks out the $s{=}3$ solution.

The contours $\C_1$ and $\C_2$ join the origin to the points
$x_1$ and $x_2$ respectively, avoiding the singularity at $x=-1$
(note that both $x_1$ and $x_2$ are less than $-1$ for this solution). 
Examining numerically the small-$|r|$ behaviour of the full TBA
equation, we found that the correct
path to take loops under $x=-1$ for $\C_1$, and over $x=-1$ for
$\C_2$.

Finally, we should settle the branches of the logarithms. 
The branches of the $L_a(\theta)$ in the TBA are such that
$L_a(+\infty)=0$ (otherwise, the integral (\ref{zeroTBAc}) would make
no sense). This translates into the branch of 
$\logca(1+x)$ being such that this function is zero
at $x=0$. From this and the way that the two contours go round the
point $x=-1$ follows the behaviour at the other ends of the contours:
\bea
\logci(1+x_1)&=&\log|1+x_1| -\imath\pi\,;\nn\\
\logcii(1+x_2)&=&\log|1+x_2| +\imath\pi\,.
\label{logip}
\eea
(Here and below, $\log x$ stands for the single-valued branch of the
logarithm, with $\im\log x\in (-\pi,\pi]\,$.)
The branch of $\logca x$ can now be determined from the stationary form
of the kink TBA equation:
\eq
\ep^{\rm kink}_a(-\infty)=\sum_{b=1}^2N_{ab}L^{\rm kink}_b(-\infty)
\label{kinktbalim}
\en
where
\eq
N_{ab}=-\frac{1}{2\pi}\iint\!\phi_{ab}(\theta)d\theta=
\left(\begin{array}{cc}1&2\\ 2&3\end{array}\right)_{ab}~.
\en
Taking the imaginary part of (\ref{kinktbalim}), and using the relations 
$\logca (x_a)=-\ep^{\rm kink}_a(-\infty)$
and $\logca(1{+}x_a)=L^{\rm kink}_a(-\infty)$, gives
\bea
\logci (x_1)&=&\log|x_1| -\imath\pi\,;\nn\\
\logcii (x_2)&=&\log|x_2| -\imath\pi\,.
\label{logxb}
\eea
Now $c(0)$ can be calculated, using the strategy advocated in 
ref.~\cite{KPa}. First deform
$\C_1$ and $\C_2$ to run along the negative real $x$ axis, save for a
small semicircle below $-1$ for $\C_1$, and a small semicircle above
$-1$ for $\C_2$. The contours can then be split into segments and the
integrands reexpressed in terms of the single-valued logarithm $\log
x$, always with positive real arguments. 
For $L_+(\C_1)$ this goes as follows. We have 
\[
2L_+(\C_1)=A-B
\]
with
\bea
A&=&\int_{\C_1}\frac{\logci(1{+}x)}{x} dx \nn\\
 &=&\int_0^{-1}\frac{\log(1{+}x)}{x} dx
  +\int_{-1}^{x_1}\frac{\logci(1{+}x)}{x} dx\nn\\
 &=&\int_0^1\frac{\log(1{-}x)}{x} dx
  +\int_{-1}^{x_1}\frac{\log(-1{-}x)}{x} dx
  -\int_{-1}^{x_1}\frac{\imath\pi}{x} dx\nn\\
 &=&\int_0^1\frac{\log(1{-}x)}{x} dx
  +\int_{0}^{-x_1-1}\frac{\log x}{x{+}1} dx
  -\imath\pi\log(-x_1) \nn
\eea
and
\bea
B&=&\int_0^{x_1}\frac{\logci (x)}{1{+}x}dx \nn\\
 &=&\int_0^{x_1}\frac{\log(-x)}{1{+}x}dx
   -\int_0^{x_1}\frac{\imath\pi}{1{+}x}dx \nn\\
 &=&-\int_0^1\frac{\log x}{1{-}x}dx
   -\int_0^{-x_1-1}\frac{\log(1{+}x)}{x}dx
   -\pi^2-\imath\pi\log(-1{-}x_1)\,. \nn
\eea
(In the second calculation, any integrals from $0$ to $x_1$
should be understood as including the small semicircle below $-1$.)
Collecting the pieces together and dividing by $2$, the final result for
$L_+(\C_1)$ is
\[
L_+(\C_1)=-\rdilog(1)-\rdilog(1+1/x_1)+\pi^2+\imath\pi\log
\left[(1{+}x_1)/x_1\right]~.
\]
The calculation of $L_+(\C_2)$ is similar, save for the switch to
$\logcii$ and the fact that
the semicircular part of the contour must now go above the point
$x=-1$ instead of below it, causing $+\pi^2$ to be replaced by $-\pi^2$.
Adding everything together,
\[
\sum_{a=1}^2L_+(\C_a)=-\sum_{a=1}^2(\rdilog(1)+\rdilog(1{+}1/x_a))
+\frac{\imath\pi}{2}\log\left[(1{+}x_1)(1{+}x_2)x_2/x_1\right]~.
\]
The stationary Y-system, satisfied by the $1/x_a$, can be used to show
that the argument of the logarithm
is equal to $1$, and hence the imaginary part of the
sum vanishes. For the rest, substituting $x_a=1/y^{(3)}_a$, and using
$\rdilog(1)=\pi^2/6$, and the sum rule (\ref{ruletwo}), gives
\[
c(0)=-20/7\,,
\]
which is the expected answer.

\subsection{Singular lines and branch points}
\label{singlsec}
The basic TBA continues to apply so long as none of the
singularities of $L_1$ or $L_2$ touch the real axis.
To put this consideration into a more general setting, imagine that
the full Riemann surface covering the complex $r$ plane
is marked with a set of 
lines, along each of which a given singularity in $L_1$ or $L_2$ has a
vanishing imaginary part.
As one of these `singular lines' is crossed,
singularities change their status from inactive to
active or vice versa, and a transition is induced in the TBA
equations. 
There is a distinction to be made between the two 
different types of singularity that the $L_a(\theta)$ can
have, one stemming from $z_a(\theta)=0$, the other from
$Y_a(\theta)=0$. Transitions associated with the zeroes of
the $z_a(\theta)$
always result genuinely different TBA equations, and we will refer to
the corresponding singular lines as being of `type I'. The same
turns out not to be true of
the transitions associated with zeroes of the $Y_a(\theta)$,
and these singular lines will be called
`type II'. Further discussion of this point will be delayed until
section~\ref{typetwosec}, when a concrete example will be examined.

It is important to realise that these singular lines are 
associated with transitions at the level of the TBA equations,
and {\it not} with singular behaviour of the $Y$'s as functions
of $r$, still less with singular behaviour of the scaling functions.
For example, if we had chosen to set up the TBA equations using a 
contour other than the real axis for all of the integrations, then
the set of lines across which equations changed from one
form to another would have been differently placed.
Nevertheless, the singular lines do
provide an effective way to map the full Riemann surface, and
they are also very useful as a means to organise the various
generalised TBA equations.
However, even
for a model as simple as the one under consideration here, it is
by no means an easy task to find their pattern.
The main problem is that at a point on the
Riemann surface near to a singular line, a singularity in $L_1$ or 
$L_2$ will inevitably be near to the real $\theta$ axis, and 
this tends to destabilise the 
iteration schemes used to solve the equations.
This makes it rather hard to disentangle the genuine boundary of 
applicability of a given equation from its `region of stability', the
latter region depending on the particular numerical method employed
and therefore having no intrinsic interest. One technique to
get around this difficulty is to locate the singularity positions with
high accuracy in those regions where iteration does work well, and then
to extrapolate these positions into the `difficult' regions around the
singular lines. Especially in situations where data can be 
obtained on both sides of a given line, this allows its location
to be found with a fair degree of confidence. However
the method is distinctly tedious to implement, and there is 
always the possibility that a crucial singularity has been missed. 
In some cases, it is possible to do much better. The key idea is 
to make use of the shifted conjugation property (\ref{yconj}). If a 
singularity in $L_a(r,\theta)$
crosses the real axis, the corresponding point for
$\tilde L_a(\tilde r,\theta)$ will be crossing the line 
$\im\theta=7\pi/10$,
a harmless piece of behaviour from the point of view
of the equations for $\tilde\ep_a$. This makes it
possible to find the functions $Y_a$ at many points where the direct
attempt at numerical solution of the relevant TBA
fails, simply by solving the equations for $\tilde Y_a$ instead.

This remark can be put to immediate use, to resolve an apparent paradox
about the behaviour of the solutions discussed so far in the
neighbourhood of the point $\rho{=}\rho_0$
on the positive-$\lambda$ line.
Recall that along the initial segment
of the positive-$\lambda$ line, the singularities of the $L_a(\theta)$ 
nearest to the real axis were at $\theta=\pm\theta^{(j)}_a$, and all had
imaginary parts equal to $\pm 3\pi/20$. At first sight this is puzzling:
in the case of the Ising model the branch points in the scaling
functions can all be associated with the `pinching' of the real $\theta$
axis by pairs of colliding singularities (see \cite{DTa} for a discussion), 
and yet
here, as we approach the point $\rho{=}\rho_0$ at which there is
a branch point in $c(r)$, the singularities in
$L_1$ and $L_2$ show no inclination at all to pinch the real axis. 
The resolution is rather appealing: the interacting nature of this model
as compared to Ising, reflected in the fact that a TBA equation must
be solved before the pseudoenergies can be found, serves to `desingularise'
the pinch. Instead, in the immediate vicinity of $\rho{=}\rho_0$,
the branching of $c(r)$ is due entirely to the multivalued nature of the
solutions to the basic TBA equation. Get a little further away, and 
singularities start to cross the real axis as the point $\rho{=}\rho_0$
is encircled, making the analysis similar to that of a pinch singularity.
This fact, coupled to the severe instability of the `excited' solution near
to the branch point, makes it easy to confuse the situation with a genuine
pinch singularity. However, with the help of the conjugation
trick just described, the numerical problems can be circumvented and the
ambiguity resolved. We first observe that as $\rho$ increases beyond 
$\rho_0$, the functions $Y_a(\theta)$ found from 
ground-state solution to the basic TBA cease to be
self-conjugate. Hence the conjugated functions
$\tilde Y_a(\theta)$ start to differ from the
$Y_a(\theta)$, and provide alternative solutions
to some set of TBA equations. 
For $\rho_0<\rho<2.8$, none of the singularities in $\tilde
L_1(\theta)$ and $\tilde L_2(\theta)$ have crossed
the real $\theta$ axis, and so the
$\tilde\ep_a(\theta)$ continue to solve the
basic TBA equation, rather than one of its generalisations. 
We used these functions in (\ref{zeroTBAc})
to find a scaling function
in the segment $\rho_0<\rho<2.8$ of the positive-$\lambda$ line, 
with results which matched perfectly with the energy level
missed by the ground-state solution. (Note that while the two levels
appear to sit on top of each other for $\rho>\rho_0$ on figure
\ref{TCSA3}, this is an illusion caused by the equality of their real
parts -- the imaginary parts are different.) Moving beyond $\rho=2.8$,
pairs of singularities in the conjugated functions
(first in $\tilde L_2$, and then, at around $\rho=4.2$,
in $\tilde L_1$) cross the real axis and
the basic TBA together with the conjugation trick provides us, for 
free, with solutions to a couple of generalised TBA equations. 

In the ultraviolet limit, that is in
those regions of the Riemann surface covering the neighbourhood of the
point $r{=}0$, another simple observation constrains
the pattern of singular lines rather strongly.
The idea is to exploit the control over the ultraviolet singularity
positions given to us by the splitting into kink systems, and the
consequent formula (\ref{fullc}).
This information can be used in the following way.
Since $\log(1/r)=\log(1/|r|)-\imath\arg(r)$, 
changing the modulus of $r$, once the kink systems have formed, does not
affect the imaginary parts of the singularity locations. 
However, a change in the value of $\arg(r)$ does have an
effect, and may result in a singularity
hitting the real $\theta$ axis. Once this has happened, varying  $|r|$ again,
while holding $\arg(r)$ fixed,
just slides this singularity along the real axis
without shifting its imaginary part away
from zero.  Hence the 
asymptotic pattern of the singular lines near to the origin is 
a collection of rays emanating from $r{=}0$. Their directions follow
from (\ref{fullc}): for the type~I lines, associated with
zeroes of $z_a(\theta)$, they are
\eq
\arg(r)=\pm\lf(\im c^{(j)}_a\pm\frac{\pi}{5}\ri)
\label{zlines}
\en
while for the type~II lines, associated with zeroes of $Y_a(\theta)$,
the relevant directions are
\eq
\arg(r)=\pm\im c^{(j)}_a~.
\label{ylines}
\en
For the ground state pseudoenergies solving the basic TBA equation,
all of the $c^{(j)}_a$ have imaginary parts equal to $7\pi/10$.
It follows that there
are no singular lines approaching the origin
in the sector $-\pi/2\le\arg(r)<\pi/2$
of the first sheet of the Riemann surface.
However for the excited solution, which was initially discussed only on the
positive-$\lambda$ line $\arg(r)=7\pi/20$, we saw that
$\im c^{(0)}_2$ vanished, and so (\ref{zlines}) implies that a type~I
singular line approaches the origin on the second sheet
along $\arg(r)=\pi/5$. As this line is traversed, a zero
of $z_2(\theta)$ crosses the real axis and the pseudoenergies
start to be determined by the first generalised TBA equation, to be
discussed shortly. In fact we already know a
kink solution to this equation:
it is just the excited kink solution to the basic TBA, shifted
yet further in the imaginary $\theta$ direction as appropriate 
for the diminishing
value of $\arg(r)$. To find out what happens to this solution 
as the ultraviolet region is left, we now turn to a direct analysis of the
generalised TBA equation that it solves.

\subsection{One pair of active singularities}
\label{oneTBAsec}
As $r$ crosses the type~I singular line on the second sheet,
the singularities in $L_2(\theta)$ at $\pm\theta^{(0)}_2$
cross the real axis, the imaginary part of $\theta^{(0)}_2$
becoming for the first time positive.
Contours in integrals involving $L_2(\theta)$ must be
distorted, and returning them to the real axis
induces extra residue terms in the equations. Just as 
described in ref.~\cite{DTa} for the SLYM, 
these residues can be found via an integration by parts. 
Equations (\ref{zeroTBA}) and (\ref{zeroTBAc}) become
\eq
\ep_a(\theta)=m_a r\cosh\theta+\logc{S_{a2}(\theta-\theta_2^{(0)})\over
S_{a2}(\theta+\theta_2^{(0)})}-\sum_{b=1}^{2} \phi_{ab}{*}L_b(\theta)
\label{oneTBA}  
\en
and
\eq
c(r)= {12r \over\pi}\imath\, m_2 \sinh \theta_2^{(0)}
 +{3 \over \pi^2} \sum_{a=1}^{2} \iintd\,
 m_a r\cosh\theta L_a(\theta)
\label{oneTBAc}
\en
respectively. 
These equations must be handled with some care. A continuous branch of
the logarithm, signalled by $\logc$, is implied.
We set
\eq
\logc S_{ab}(+\infty)=0\quad;\quad L_b(+\infty)=0\,,
\label{brch}
\en
and then understand $\logc(S/S)$ to be continuous as $\theta$ varies
along the real axis. For the moment we shall also assume 
$0<\im\theta^{(0)}_2<\pi/5$; the effects of $\theta^{(0)}_2$ straying
beyond this strip will be examined later.
Using $\phi_{ab}=-\imath\frac{\partial}{\partial\theta}
\log S_{ab}(\theta)$, we have
\[
\left[\logc S_{ab}\right]_{-\infty}^{\infty}
=\imath\iint\!\phi_{ab}(\theta)d\theta\equiv -2\pi\imath N_{ab}
=-2\pi\imath \left(\begin{array}{cc}1&2\\ 2&3\end{array}\right)_{ab}~.
\]
Given the branch choice made above, and also the symmetry of
$S_{ab}(\theta)$, this implies that
\eq
\logc S_{ab}(0)=\pi\imath N_{ab}\quad;\quad
\logc S_{ab}(-\infty)=2\pi\imath N_{ab}~.
\en
Substituting $\theta=\theta^{(0)}_2$ into (\ref{oneTBA}),
and imposing $\ep_2(\theta^{(0)}_2)=\imath\pi$ to ensure that
$z_2(\theta)$ vanishes at that point, then gives the final ingredient:
\eq
0=m_2 r\cosh\theta_2^{(0)}+\pi\imath(N_{22}{-}1) -\logc
S_{22}(2\theta_2^{(0)})-\sum_{b=1}^{2} \phi_{2b}{*}L_b(\theta_2^{(0)})~.
\label{oneTBAt}  
\en
The term $\pi\imath N_{22}$ comes
from the consistent application of the branch choice (\ref{brch}) to
$\logc S_{22}(0)$.
Notice that it is equal to $3\pi\imath$ and {\it not} the $\pi\imath$
that might at first sight have been expected -- the importance of this
will become apparent during the calculation of $c(0)$ below.
The remaining logarithm must also be taken continuously, starting from
the branch found when $\theta^{(0)}_2$ crosses the axis.
This is not completely trivial, as $2\theta^{(0)}_2$ may ultimately
leave the strip $|\im 2\theta^{(0)}_2|<\pi/5$, making it necessary
to keep track of the sense in which the pole in
$S_{22}(\theta)$ at $\theta=\imath\pi/5$ is encircled.

All of this makes the analytic treatment a little tricky.
Perhaps surprisingly, it is much less of a problem for the numerical
solution of the equations. We used the following method:
start by choosing an initial value for
$\theta^{(0)}_2$, and then iterate (\ref{oneTBA}) until a 
solution $\ep_a(\theta|\theta^{(0)}_2)$ is found. 
At this stage the first potential difficulty arises:
unless otherwise instructed,
Fortran will always choose the principal (discontinuous) branch for
the logarithm during these iterations, $-\pi<\im\log(S/S)\le\pi$. But any
$2\pi\imath$ discontinuities that this induces in the resulting
functions $\ep_a(\theta|\theta^{(0)}_2)$ 
leave the values of the $L_a=\log(1{+}\exp(-\ep_a))$ unchanged, both in
(\ref{oneTBA}) and (\ref{oneTBAc}), and so this problem can be 
ignored. 
The functions  $\ep_a(\theta|\theta^{(0)}_j)$ are then inserted
into (\ref{oneTBAt}); the error in  the initial
$\theta^{(0)}_2$  will be reflected in the degree to which this
equation fails to hold. Again, one might worry that Fortran will spoil
things by picking the wrong branch for the logarithm; this time, we
sidestepped the problem by exponentiating the equation. Once this has
been done, an adeptly-chosen portion of $S_{22}(2\theta^{(0)}_2)$ can
be taken over to the left-hand side, and then inverted to obtain an
improved estimate for $\theta^{(0)}_2$. The whole procedure is then
iterated until a stable solution is found. 
In this way we were able to map out large areas of the 
complex $r$ plane, but not, as it happens, a segment $0<r<r_c$
of the real axis. The reason for this difficulty will be explained
later, but before that we will discuss the infrared limit.

With a little practice it is not too hard to guess how the solutions
will behave in this regime, but just to be sure we
tracked a solution and its singularities as $r$ varied
along the line $r=2\imath+(1-0.5\imath)t$, using the basic TBA 
(\ref{zeroTBA}) for $0.1<t<1.5$, and the generalisation 
(\ref{oneTBA}), (\ref{oneTBAt}) for $1.5<t<4$. We found that the
singularity positions are
indeed smoothly continued by the new equations. When $r$ finally
reaches the real axis, the singularity positions regain the 
standard conjugation symmetry under $\theta\rightarrow\theta^*$, as
oppose to the shifted conjugation symmetry under $\theta\rightarrow\tilde
\theta$ associated with the positive-$\lambda$ line. However the way
in which the symmetry is realised is rather different from the situation
before any
continuation, when the basic TBA equation applied and singularity
pattern was as shown in figure~\ref{epsA}.
Then, all of the $\beta^{(j)}_a$ lay on a fixed line of the mapping
$\theta\rightarrow\theta^*$, namely $\im\theta=7\pi/10$. (Recall that 
all functions in the game are $7\pi\imath/5$-periodic, so this is 
indeed a fixed line.) This time, $\beta^{(0)}_1$ and $\beta^{(0)}_2$
are instead swapped with their negatives under the conjugation;
they are therefore both purely imaginary.
\begin{figure}[tbh]
\hskip 1.3cm
\epsfxsize=10.5cm
\epsfbox{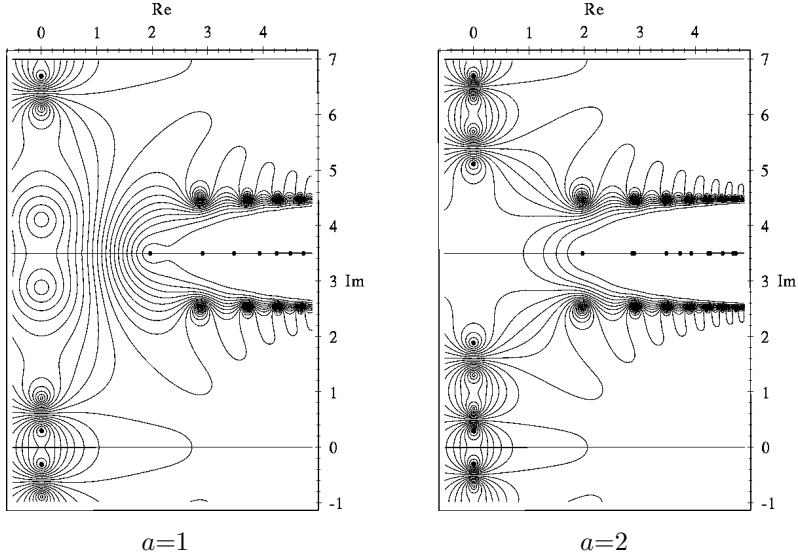}
\vskip 1pt
{\small\noindent\hfil$a{=}1$\qquad\qquad\hfil\qquad$a{=}2$\hfil\break}
\caption{  \leftskip=.5cm {\protect \small  
Contour plots of $|z_a(\theta)|/(1{+}|z_a(\theta)|)$ in the complex
$\theta$ plane, for the first generalised
TBA equation with $r{=}3$.
Labelling as in figure~\epsAnum.}
\rightskip=0.5cm }  
\label{epsE}
\end{figure}
Figure~\ref{epsE} shows the functions at $r=3$; note how, in
contrast to the earlier pictures, the zeroes of $Y_1$ and
$Y_2$ at $\beta^{(0)}_1$ and $\beta^{(0)}_2$ 
(signalled by the dots $\bullet$ on the imaginary axis) 
are clear of the fixed lines of
$\theta\rightarrow\theta^*$ and $\theta\rightarrow\tilde\theta$.

Once these qualitative features have been understood, it is possible
to extract some exact asymptotics.
On the real axis with $r$ larger than $r_c$, $\theta^{(0)}_2(r)$ lies on
the segment of the imaginary $\theta$ axis between $\imath\pi/10$ and
$\imath\pi/5$. Furthermore, during the continuation from the
ultraviolet end of the positive-$\lambda$ line, $\theta^{(0)}_2$
reaches this segment from large
positive values with no monodromy around the
point $\imath\pi/10$. Combining this information with the form of
$S_{22}(\theta)$ shows that, for real $r>r_c\,$,
\[
\logc S_{22}(2\theta^{(0)}_2(r))=
\log S_{22}(2\theta^{(0)}_2(r))+
2\pi\imath~.
\]
Equation (\ref{oneTBAt}) therefore becomes
\eq
0=m_2 r\cosh\theta_2^{(0)}-\log
S_{22}(2\theta_2^{(0)})-\sum_{b=1}^{2} \phi_{2b}{*}L_b(\theta_2^{(0)})~.
\label{oneTBAtt}  
\en
Since $S_{22}(2\theta)$ is real and positive for imaginary $\theta$
between $\imath\pi/10$ and $\imath\pi/5$,
all three terms in this equation are real.
As $r$ grows, the first two terms 
must cancel, as the convolution term tends to zero. 
To achieve this, $\theta^{(0)}_2$ approaches the
limiting value of $\imath \pi/10$, at which $S_{22}(2\theta^{(0)}_2)$
has a pole. The leading behaviour of $c(r)$ follows immediately:
substituting $\theta^{(0)}_2=\imath\pi/10$ into (\ref{oneTBAc}) and using
the relation
\eq
m_1=  2 m_2 \sin({\pi\over 10})
\label{mass1}
\en
gives $c(r)\sim -6m_1r/\pi$, or $E(\lambda,R)\sim E_{\rm bulk}(\lambda,R)
+M_1$. This confirms that the state under discussion is a one-particle
state, with mass gap $M_1$. Since it also has zero momentum (the
total momentum of a level is
conserved under analytic continuation), it must be
the first line above the ground state in figure~\ref{TCSA2}. 
As a further check, a couple of corrections can be worked out. These
should reproduce the formulae for finite-volume mass shifts found in 
refs.\cite{La,KMd}, and summarised in appendix~\ref{ftres}. 
The more precise cancellation of the first two terms 
in (\ref{oneTBAtt}) requires
\eq
\theta_2^{(0)}(r)\sim\imath\lf(\pi/10+ 
\tan(\frac{3\pi}{10})\tan(\frac{2\pi}{5})^2
e^{- m_2 r 
\cos(\pi/10)}\ri)\,.
\en
This improved estimate for $\theta^{(0)}_2$ 
gives one correction, the so-called $\mu$-term,
when substituted into (\ref{oneTBAc}).
A second comes on expanding $L_a(\theta)$ in (\ref{oneTBAc}),
using the leading behaviour of $\ep_a(\theta)$ obtained from the first 
two terms on the RHS of (\ref{oneTBA}). For this correction we can set
$\theta^{(0)}_2=\imath\pi/10$, and then use the identity
\eq
\frac{S_{a2}(\theta+\imath{\pi\over 10})}{S_{a2}(\theta-\imath{\pi\over 10})}
=S_{a1}(\theta+\imath {\pi \over 2})~.
\label{ss1}
\en
Putting these pieces together, the 
two leading IR corrections are
\bea
c(r)\!\! &\sim& \!\!{-6r\over\pi}\Bigl(m_1+2m_2\cos(\frac{\pi}{10})
\tan(\frac{3\pi}{10})\tan(\frac{2\pi}{5})^2
e^{-m_2r\cot(\pi/10)} \nn \\
 &&\qquad\qquad\!\!{}-{1\over 2\pi}\sum_{a=1}^{2}\iintd\,
 m_a \cosh\theta~ S_{1a}(\theta+{\imath \pi\over 2})e^{-rm_a\!\cosh 
\theta }\Bigr)~.\qquad~~~
\eea
These agree with the expected results, as given in 
ref.~\cite{KMd}.

\begin{figure}[tbh]
\hskip 1.3cm
\epsfxsize=10.5cm
\epsfbox{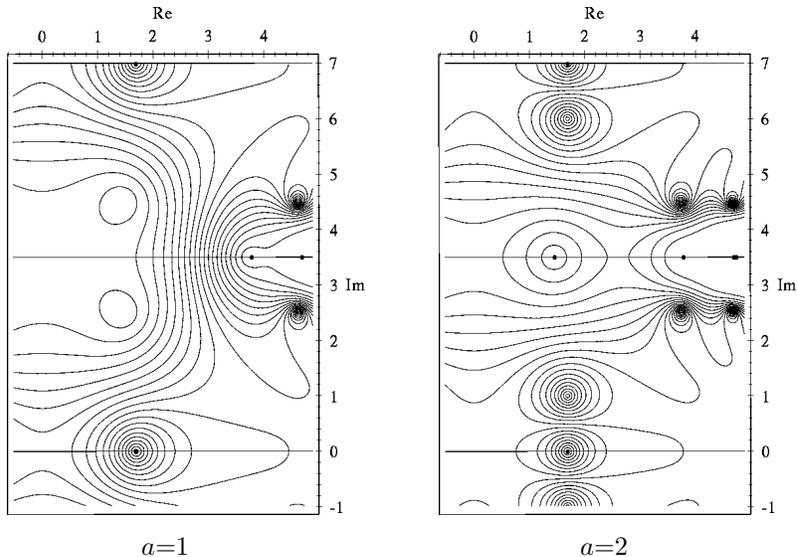}
\vskip 1pt
{\small\noindent\hfil$a{=}1$\qquad\qquad\hfil\qquad$a{=}2$\hfil\break}
\caption{  \leftskip=.5cm {\protect \small  
Contour plots of $|z_a(\theta)|/(1{+}|z_a(\theta)|)$ in the complex
$\theta$ plane, for the generalised TBA equation with $r{=}1$.
Labelling as in figure~\epsAnum.}
\rightskip=0.5cm }  
\label{epsC}
\end{figure}

In the opposite, ultraviolet limit the singularity pattern is rather
different.
As remarked earlier,
once the limiting form of the excited solution to
the basic TBA on the
positive-$\lambda$ line near to $r=0$ is known,
the fact that the only $r$-dependence of the kink solutions
is in the $\pm\log(1/r)$ anchoring points for the left and right
kink systems allows the solution to be continued back
to the real axis, thereby obtaining
a kink solution to
the generalised TBA equations (\ref{oneTBA}) and (\ref{oneTBAt}).
Recall from just after (\ref{fullc}) that for the excited solution to the
basic TBA
all of the $c^{(j)}_a$ had imaginary parts equal to $7\pi/10$ apart
from $c^{(0)}_2$, which had an imaginary part equal to zero.
Substituting into (\ref{fullc}) shows
that for real $r$ the kink limit of this
solution to the generalised TBA equation
(\ref{oneTBA}) has
the imaginary parts of all of the $\beta^{(j)}_a(r)$ 
equal to $(\pm)7\pi/10$, apart from 
$\beta^{(0)}_2(r)$, for which $\im\beta^{(0)}_2(r)=0$. 
This also holds
before the kink limit is taken, since when
$r$ is real, the singularity pattern must be symmetrical about the
real $\theta$-axis. For a zero of $Y$ to move away from a fixed line
of $\theta\rightarrow\theta^*$,
there would have to be a corresponding zero, with equal
real part, moving
in the opposite direction. This is only possible once the left
and right kink systems have met. (Numerically we found that this
happens at the previously-mentioned point $r=r_c\approx 2.6646510318(2)\,$.)
Figure~\ref{epsC} shows the solution at $r{=}1$.
Notice that the pattern of singularities is indeed symmetrical under
$\theta\rightarrow\theta^*$, but in a different way from the infrared
solution shown in figure~\ref{epsE}. The situation is
similar to that observed for the first excited state of the SLYM in
refs.~\cite{BLZa,DTa}: as $r$ increases
from the ultraviolet,
the two kink systems come into contact and the TBA profits from the
extra possibilities that this offers for satisfying the
conjugation symmetry, and rearranges the singularity pattern into
a form more appropriate for the infrared limit. 
For real $r<r_c$ we have
$\re\beta^{(0)}_2\neq 0$ and $\im\beta^{(0)}_2=0$, while for real
$r>r_c$,
$\re\beta^{(0)}_2=0$ and $\im\beta^{(0)}_2\in (-\pi/5,0)$. (There is
also a transition associated with $\beta^{(0)}_1$, when it moves from the
line $\im\beta^{(0)}_1=7\pi/10$ onto the imaginary axis. This happens at
$r\approx 1.5$, but since the associated 
singularities are inactive, the equations do not change.)
The transition at $r{=}r_c$ will be analysed further below, but first
we will discuss the
ultraviolet behaviour of the equations, where the singularities 
follow the pattern predicted by the excited solution to the
basic TBA equation.

There is a new complication in the ultraviolet regime when $r$ is
real. Since $\beta^{(0)}_2$ is real, 
(\ref{y1tab}) implies that singularities at $\pm\beta^{(0)}_2$ in
$L_1$ and $L_2$ are sitting exactly on the integration contour.
For analytic work, the most convenient approach seems to be to take 
these singularities into account via small
distortions of all contours of integration, below the real axis near
to $\theta=-\beta^{(0)}_2$, and above the real axis near to
$\theta=\beta^{(0)}_2$. This is rather like the
usual `$\imath\epsilon$' prescription in quantum field theory, and 
amounts to giving $r$ an infinitesimal positive-imaginary part.
Here, the justification for the prescription comes from
the analytic continuation idea that has been fundamental from the
outset. (In ref.~\cite{DTa}, we instead rewrote the equations
using principal values for all integrals. Whilst this puts the
equations into an appealingly symmetrical form, the branches of the
logarithms are hard to disentangle, and the treatment below seems to
be more promising for systematic work.)

To verify that the value of $c(0)$ given by (\ref{oneTBAc}) and the
solution to the generalised TBA equation is the same as came via
the excited solution to the basic TBA, we start by finding the kink
form of the generalised TBA system. From (\ref{fullc}) we have, as
$r\rightarrow 0$,
\[
\beta^{(0)}_2(r)\sim \log(1/r)-c^{(0)}_2
\]
where, for the solution under discussion,
$c^{(0)}_2$ is a real
constant. Hence
\eq
\theta^{(0)}_2\sim \log(1/r)-c^{(0)}_2+\frac{\imath\pi}{5}~.
\en
The $\imath\epsilon$ prescription for $r$ means that $\theta^{(0)}_2$
is still (just) inside the strip $0<\im\theta^{(0)}_2<\pi/5$. We can
take this into account by setting $r=\rho e^{\imath\epsilon}$, with
$\rho$ real and $\epsilon$ understood to be vanishingly small and
positive. The kink
form of equations (\ref{oneTBA}), (\ref{oneTBAt}) and (\ref{oneTBAc})
follows on shifting $\theta$ to $\theta{+}\log(1/\rho)$, defining
$\ep^{\rm kink}_a(\theta)=\ep_a(\theta{+}\log(1/\rho))$, employing the
limiting form of $\theta^{(0)}_2(r)$, and finally taking the
$\rho\rightarrow 0$ limit, dropping inessential factors of
$\imath\epsilon\,$:
\bea
&&\ep^{\rm kink}_a(\theta)=
\halft m_a re^{\theta}+\logc S_{a2}(\theta{+}c^{(0)}{-}
\fract{\imath\pi}{5}
{+}\imath\epsilon)-\sum_{b=1}^{2} \phi_{ab}{*}L^{\rm kink}_b(\theta)\,;
\qquad\quad
\label{onekinkTBA}\\
&&0=\halft m_2 e^{-c^{(0)}}\!e^{\imath\pi/5}+\pi\imath(N_{22}{-}1)
-\sum_{b=1}^{2}\phi_{2b}{*}
L^{\rm kink}_b(-c^{(0)}{+}\fract{\imath\pi}{5}{-}\imath\epsilon)\,;
\quad\label{onekinkTBAt}\\
&&c(0)={6\over\pi}\imath\, m_2 e^{c^{(0)}}\!e^{\imath\pi/5}
 +{6\over\pi^2}\sum_{a=1}^{2}\iintd\,
 \halft m_ae^{\theta}L^{\rm kink}_a(\theta)\,.
\label{onekinkTBAc}
\eea
(For the last equation, use is also made of the fact that half of the
integral contributing to $c(0)$ comes from the left-hand kink system,
and half from the right.) 
Differentiating (\ref{onekinkTBA}) with respect to $\theta$ and
substituting the resulting expression for $\halft m_ae^{\theta}$ into
(\ref{onekinkTBAc}) gives
\bea
c(0)\!&=&\!
{6\over\pi}\imath\, m_2 e^{c^{(0)}}\!e^{\imath\pi/5}
+{6\over\pi^2}\sum_{a=1}^{2}\iintd\,
\Bigl[\prtial\ep^{\rm kink}_a(\theta) \nn\\
&&\quad\qquad\qquad{~} -\imath\phi_{a2}(\theta{+}
c^{(0)}_2{-}\fract{\imath\pi}{5})+\sum^2_{b=1}\phi'_{ab}{*}
L_b^{\rm kink}(\theta)\Bigr] L^{\rm kink}_a(\theta)\,.\nn
\eea
The use of (\ref{onekinkTBAt}) simplifies this to
\bea
c(0)\!&=&\! 12(N_{22}{-}1) +
\frac{6}{\pi^2}\sum_{a=1}^2\int_{\E_a}\logea(1{+}e^{-\ep})d\ep\nn\\
&&\qquad\qquad\qquad
{} +\frac{6}{\pi^2}\sum_{a,b=1}^2\iintd\left[\phi'_{ab}{*}L_b^{\rm kink}
(\theta)\right]L_a^{\rm kink}(\theta) \nn\\
&=&\! 12(N_{22}{-}1) +
\frac{6}{\pi^2}\sum_{a=1}^2\int_{\E_a}\logea(1{+}e^{-\ep})d\ep\nn\\
&&\qquad\qquad\qquad
{} -\frac{3}{\pi^2}\sum_{a,b=1}^2
N_{ab}\left[L_a^{\rm kink}
(\theta)L_b^{\rm kink}(\theta)\right]^{\infty}_{-\infty}~. \nn
\eea
Here $\E_a$ is the contour swept out by $\ep^{\rm kink}_a$, and
$\logea$ is a continuous choice of the logarithm along this contour.
The branch choice made previously
implies that $L^{\rm kink}_b(+\infty)=0$, while for
$L_b^{\rm kink}(-\infty)$ 
the $\theta\rightarrow-\infty$ limit of (\ref{onekinkTBA}),
\eq
\ep^{\rm kink}_a(-\infty) =2\pi\imath N_{a2}+
\sum_{b=1}^2N_{ab}L_b^{\rm kink}(-\infty)\,,
\label{newminf}
\en
can be used.
Rearranging and changing variables in the integrals,
\bea
c(0)&=&12(N_{22}{-}1)+\frac{6}{\pi^2}\sum_{a=1}^2\left[
L_+(\C_a)-\pi\imath N_{a2}L^{\rm kink}_a(-\infty)\right]\nn\\
&=& -12-\frac{6\imath}{\pi}\ep_2^{\rm kink}(-\infty)+
\frac{6}{\pi^2}\sum_{a=1}^2 L_+(\C_a) \,,
\label{lastc}
\eea
with $L_+(\C_a)$ as defined in (\ref{dilog}). The contours are
a little different from the previous case: a single zero in
$z_2(\theta)$ crossed the real axis in the continuation from the
basic TBA to the new equation. With no further numerical work
we can deduce that this time both contours go below the point $x=-1$.
As a result, (\ref{logip}) is replaced by
\bea
\logci(1+x_1)&=&\log|1+x_1| -\imath\pi\,;\nn\\
\logcii(1+x_2)&=&\log|1+x_2| -\imath\pi\,.
\label{revlogip}
\eea
The branches of $\logca(x)$ should be unchanged, and indeed
substituting (\ref{revlogip}) into the imaginary part of (\ref{newminf})
shows that (\ref{logxb}) still holds. From this point on the
evaluation of the integrals runs essentially
as before, though note that this time the residue
terms do not cancel but rather reinforce each other. The result
is
\[
\sum_{a=1}^2L_+(\C_a)=-\sum_{a=1}^2(\rdilog(1)+\rdilog(1{+}1/x_a))
+\pi^2
+\frac{\imath\pi}{2}\log\left[(1{+}x_1)(1{+}x_2)/(x_1x_2)\right]~.
\]
Substituting into (\ref{lastc}), and recalling that $\ep^{\rm
kink}_2(-\infty)=-\logcii(x_2)=-\log(-x_2)+\imath\pi$, the imaginary
parts again cancel and we finally
recover $c(0)=-20/7$, just as before. Assuming that the analytic
continuations had been performed correctly, this calculation had no
option but to give the right answer. We included it anyway to show how
the generalised equations can be tackled directly, so long as
the branches of the logarithms are handled in a consistent way.

{}From a numerical point of view, there is a difficulty in
the ultraviolet regime. 
The zeroes in the $Y$'s at $\pm\beta^{(0)}_2$
entail divergences in the pseudoenergies on the real $\theta$
axis, along which the TBA equation is being solved. As a result, direct
numerical iteration of the equations, as outlined above, becomes
unstable as $r$ approaches the segment $[0,r_c]$ of the real axis.
Taking a hint
from \cite{BLZa}, a useful trick is to remove these divergences
explicitly, before proceeding to a solution of the equations.
It turns out that this can be done in a particularly nice way.
The first step is the following identity, a simple
consequence of equation~(9) of ref.~\cite{Zf}:
\eq
\phi_{ad}(\theta)=-\phi_h(\theta)\,l^{[T_2]}_{ad\phantom{b}}
+\sum_{b=1}^2 \phi_{ab}*\phi_h(\theta)\,l^{[T_2]}_{bd}
\label{identi}
\en
where $\phi_h(\theta)=h/(2\cosh\frac{h\theta}{2})$ and $h$ is here 
equal to $5$. 
(For a discussion of the $T_N$-related cases of such
identities, see ref.~\cite{RTVa}.) 
This can be integrated, using
\eq
\phi_h(\theta)=\pm\imath\prtial\logc\sigma_h(\theta\pm
\frac{\imath\pi}{h})\qquad,\quad\sigma_h(\theta)=
\tanh(\frac{h\theta}{4})~,
\label{identt}
\en
to yield
\eq
\pm\logc S_{ad}(\theta)= 
\logc\sigma_h(\theta\pm\frac{\imath\pi}{h})\,l^{[T_2]}_{ad\phantom{b}}
 - \sum_{b=1}^2\phi_{ab}{*}
\logc\sigma_h(\theta\pm\frac{\imath\pi}{h})\,l^{[T_2]}_{bd}~.
\label{lSresult}
\en
Substituting the $d{=}2$ case into
(\ref{oneTBA}), recalling that
$\beta^{(0)}_2=\theta^{(0)}_2{-}\imath\pi/h$, 
and absorbing the extra terms by defining
\eq
\hat\ep_a(\theta)= \ep_a(\theta) -
\logc\left(\sigma_h(\theta{-}\beta_2^{(0)})
\sigma_h(\theta{+}\beta_2^{(0)})\right)
\en
and 
\eq
\hat L_a(\theta)= \logc\lf( 
\sigma_h(\theta{-}\beta_2^{(0)}) 
\sigma_h(\theta{+}\beta_2^{(0)}) 
+ e^{-\hat\ep_a(\theta)} \ri)~,
\en
the excited-state TBA equation has a form identical to the original
equation for the ground state:
\eq
\hat\ep_a(\theta)=m_a r\cosh \theta-\sum_{b=1}^{2} 
\phi_{ab}{*}\hat L_b(\theta)~. 
\label{oneTBAg}  
\en
The different singularity structure is now entirely encoded in the
definitions of $\hat\ep_a$ and $\hat L_a$, which, in contrast to
$\ep_a$ and $L_a$, are not themselves singular at $\pm\beta^{(0)}_2$.
The expression for the central charge also has a formal similarity
with the original TBA: 
the integral 
\eq
{1\over 2\pi}\iintd e^{\imath  k\theta} 
\log\lf(\sigma_h(\theta{-}x{+}\imath{\pi\over h})\sigma_h(\theta{+}x{-}\imath 
{\pi \over h}) \ri) = {\imath  \sin k x \over  k \cosh(\pi k/h) }  
\en
(convergent in the region $|\im k|<h/2$), taken at $k=\imath$ and 
$x=\theta^{(0)}_2$, allows 
equation (\ref{oneTBAc}) to be rewritten as
\eq
c(r)= {3 \over \pi^2} \sum_{a=1}^{2} \iintd\,
 m_a r\cosh\theta\hat L_a(\theta)~.
\label{oneTBAgc}
\en
These equations look just the same as the original TBA system. However it
is important that they contain an extra parameter, namely the value of
$\beta^{(0)}_2$. When handling the system numerically, we can't search
for the singularity at $\beta^{(0)}_2$ -- it has been explicitly removed
-- but must instead focus on $\theta^{(0)}_2$, and impose
\eq
e^{\hat\ep_2(\theta^{(0)}_2)}=
\imath\coth\lf(\fract{5}{2}\theta^{(0)}_2+\fract{\imath\pi}{4}\ri)
e^{\ep_2(\theta^{(0)}_2)}=
-\imath\coth\lf(\fract{5}{2}\theta^{(0)}_2+\fract{\imath\pi}{4}\ri)~.
\label{oneTBAgt}
\en
This method was used to obtain table~\ref{oneTBAuv}, and also figure~\ref{epsC}.

\subsection{Type II singular lines}
\label{typetwosec}
The problems encountered at the end of the last subsection can be
traced to the presence of a type II singular line, along which a zero
of $Y_a(\theta)$ has vanishing real part. As hinted earlier, the
behaviour of the generalised TBA as such a line is crossed has some novel
features.

The first point to note is that any initial crossing of the real $\theta$ 
axis by a zero of $Y_1$ or $Y_2$ must be preceded by the crossing of 
a zero of either $z_1$ or $z_2$ -- this follows from 
(\ref{y1tab}) and (\ref{y2tab}), and
the configuration of the singularities before the continuation
begins. Therefore, just prior to the first crossing
of the zero of $Y_1$ or $Y_2$,
there will already be an active singularity in the
equations, a zero in $z_c(\theta)$, say, at $\theta^{(j)}_c$.
Without any significant loss of generality, suppose that this
singularity became active as it crossed the real $\theta$ axis from
below, so that it currently lies in the strip $0<\im\theta^{(j)}_c<\pi/5$. 
The accompanying zero or zeroes of the $Y_a$, at $\beta^{(j)}_c=
\theta^{(j)}_c{-}\imath\pi/5$, lie below the real $\theta$ axis and
at this stage are inactive. 
The generalised TBA equation for $\ep_a(\theta)$ contains a term
\eq
\logc S_{ac}(\theta-\theta^{(j)}_c)
\label{modone}
\en
which can be traced to the presence of the active singularity at
$\theta^{(j)}_c$.
There is also an additive contribution to $c(r)$, equal to
\eq
\frac{6r}{\pi}\imath m_c\sinh\theta^{(j)}_c~.
\label{modtwo}
\en
Now consider what can happen to these terms as $r$ varies. The simplest
option for change is for $\im\theta^{(j)}_c$ to become once more
negative, meaning that a type~I singular line has been crossed.
Clearly this corresponds to undoing the trapped piece of
contour, and the net effect is for the two terms (\ref{modone}) and
(\ref{modtwo}) to drop out of the equations, returning them to their
original state. Note that this change in the generalised TBA equations is
signalled by $\beta^{(j)}_c$ leaving the strip
$|\im\beta^{(j)}_c|<\pi/5$.

More subtle is the behaviour when $\im\theta^{(j)}_c$
grows beyond $\pi/5$, $\beta^{(j)}_c$ crossing the real axis but
remaining inside the strip
$|\im\beta|<\pi/5$. This corresponds to the crossing of a type~II singular
line. (We already discussed a case with
$\im\theta^{(j)}_c$ exactly equal to $\pi/5$, found when $r$ hits the segment
$0<r<r_c$ of the real axis from above. The question currently being
addressed would arise if $r$ was then continued
beyond this segment into the lower half plane.) 
As $\im\theta^{(j)}_c$ passes $\pi/5$, the singularities in the
$L_a(\theta)$ at $\beta^{(j)}_c$ cross the real
axis. Further terms must be added to the equations, which are
conveniently summarised using the incidence matrix of the $T_2$ diagram,
$l_{cb}^{[T_2]}$. The terms (\ref{modone}) and (\ref{modtwo})
become
\eq
\logc S_{ac}(\theta-\theta^{(j)}_c)
-\sum_{b=1}^2l^{[T_2]}_{cb}\logc S_{ab}(\theta-(\theta^{(j)}_c{-}
\imath\pi/5))
\label{newmodone}
\en
and
\eq
\frac{6r}{\pi}\imath m_c\sinh\theta^{(j)}_c-
\frac{6r}{\pi}\imath\sum_{b=1}^2l^{[T_2]}_{cb}
m_b\sinh(\theta^{(j)}_c{-}\imath\pi/5)
\label{newmodtwo}
\en
respectively. (The minus signs are there because the new singularities
are zeroes of $Y_b(\theta)$, rather than $z_c(\theta)$.)
The situation seems to have become more complicated, but
is helped by the following pair of identities:
\[
S_{ac}(\theta-\imath\pi/5)S_{ac}(\theta+\imath\pi/5)=
\prod^2_{b=1}S_{ab}(\theta)^{l_{cb}^{[T_2]}}~;
\]
\[
m_c\left[\sinh(\theta{-}\imath\pi/5)+
\sinh(\theta{+}\imath\pi/5)\right]=\sum_{b=1}^2l_{cb}^{[T_2]}
m_b\sinh\theta~.
\]
Defining $\theta^{(j')}_c=\theta^{(j)}_c-2\imath\pi/5$, 
the terms (\ref{newmodone}) and (\ref{newmodtwo}) simplify
to
\eq
-\logc S_{ac}(\theta-\theta^{(j')}_c)
\label{finalmodone}
\en
and
\eq
-\frac{6r}{\pi}\imath m_c\sinh\theta^{(j')}_c~.
\label{finalmodtwo}
\en
These are exactly the extra terms that would have been found had a
singularity in $L_c(\theta)$, a zero of $z_c(\theta)$, crossed the
real $\theta$ axis, though from above rather than from
below. From (\ref{y1tab}) or
(\ref{y2tab}), there is indeed a zero of $z_c(\theta)$ at
$\theta^{(j')}_c$, so this reinterpretation of the equation is
consistent with the singularity pattern actually found. 

It is worth pausing to consider what has happened. The situation under
discussion was the crossing of a type~II singular line,
and we have shown that it does not
lead to a fundamentally different equation, but just a relocation of
an already-active singularity, 
back into the strip $|\im\theta|<\pi/5$.
This is in contrast to the
crossing of a type~I singular line,
which always either adds or subtracts a singularity from the active list.
It can now be
deduced that any equation obtained from the ground-state TBA system
by analytic continuation in $r$
can be rewritten in such a way that the only active singularities 
correspond to zeroes of $z_1$ or $z_2$
in the strip $|\im\theta|<\pi/5$. Furthermore, in the rewritten
equations a zero of $z_c(\theta)$ in this strip
will be active if and only if $\beta^{(j)}_c$, the position
of the  accompanying zero of
$Y_1(\theta)$ and/or $Y_2(\theta)$, satisfies $|\im\beta^{(j)}_c|<\pi/5$. 
This observation provides
the bridge between our approach and the methods advocated in
refs.~\cite{KPa,BLZa}.

Finally, we should examine the behaviour of the equations 
for the first excited state near to
$r{=}r_c$.
Recall that in the far infrared, $\theta^{(0)}_2(r)$
approached $\imath\pi/10$ from above. 
As $r$ decreases, $\theta^{(0)}_2$ grows and finally, at $r{=}r_c$,
it arrives at $\imath\pi/5$. This is necessary for there to
be a continuous transition to the singularity pattern observed when
$r$ is less than $r_c$. 
There are also inactive singularities associated with zeroes of $Y_1$ and
$Y_2$, located at $\pm\beta^{(0)}_2=\pm(\theta^{(0)}_2{-}\imath\pi/5)$,
and as $r$ approaches $r_c$,
these singularities start to pinch the real axis. 
This poses a mild puzzle, the exact converse of that discussed 
in section~\ref{singlsec}:
there, a branch point of $c(r)$ was present and
yet there was no evidence of an integration contour being pinched, while
here, contours are being pinched 
and yet (because $r$ is real) we do not expect
there to be a branch point.
The absence of this branch point can be understood if we use
the standard technique for analysing a pinch
singularity, as described, for example, in chapter~2 of \cite{ELOP}.
This involves continuing $r$ in a complete circle around the 
potentially singular
point, and then comparing the result with the uncontinued situation in
order to find the discontinuity over the putative branch cut.

Near to $r_c$, we found the following dependence of 
$\theta^{(0)}_2$ on $r$:
\eq
\theta_2^{(0)}(r)=\imath{\pi\over 5}+ B(r_c{-}r)^{1/2} +O(r_0{-}r)
\label{thetaroot}
\en
with $B=0.3763188677(2)$.
This gives a good control over the positions of the singularities,
allowing their changing status to be followed as $r$ traces out a 
small circle about $r_c$.
At first sight the situation after continuation looks very different
from that before: the active singularities at
$\pm \theta_2^{(0)}$ have left the strip $|\im\theta|<\pi/5$, their 
former positions now being
occupied by a pair of inactive singularities, whilst
the two singularities at $\pm \beta^{(0)}_2$
have swapped over and become active. 
However, while the assignments of singularities
as active or inactive have changed, their overall pattern has not.
Even better, a moment's thought shows that what has happened is just as
would be expected, given that the
type~II singular line running from $0$ to $r_c$ 
along the real axis was crossed exactly once during the continuation.  
The earlier discussion of
such situations applies, and if the equations are rewritten as described
above, they
return to exactly the form that they had before
the continuation was performed. Hence the discontinuity
is zero, and there is no branch cut
after all. The functions $Y_a(r,\theta)$
are single-valued in the neighbourhood of $r{=}r_c$, despite the square root
in (\ref{thetaroot}), which merely shows that the labelling 
of the singularities is not single-valued as $r_c$ is encircled -- it
does not of itself require that the $Y$'s should be multiple-valued.

One particular consequence of this discussion is that the
type~II singular line actually terminates at
$r{=}r_c$.
This behaviour is special to
the lines associated with zeroes of the $Y$'s, and is possible because
the number of zeroes of the $Y_a(\theta)$ in the strip 
$|\im\theta|<\pi/5$ does not change as such
lines are crossed. By contrast, type~I lines separate regions of 
validity of fundamentally different TBA equations, and so they
cannot just stop at isolated values of $r$. In fact, they
always seem to be anchored on some covering point of $r{=}0$, although
we cannot absolutely rule out other closed loops elsewhere on
the Riemann surface.

%
\subsection{On to the third sheet}
\label{thirdsheetsec}
The third sheet can be reached by analytic
continuation around either $B$ or $\tilde B$ on figure~\ref{surface}.
The point $\tilde B$ lies in a part of the Riemann surface described by the
basic TBA equations
(\ref{zeroTBA}) and (\ref{zeroTBAc}), and
continuing around $\tilde B$ in an anticlockwise sense we find a third
set of solutions to this system. If $r$ is now returned to the origin
keeping $\arg(r)$ greater than approximately $2.11\pi/5$, no singularities 
cross the real
$\theta$ axis and a new kink solution to the basic TBA is uncovered.
The value of $c(0)$ on the third sheet can now be calculated much as
before, using the new forms of the 
contours $\C_1$ and $\C_2$ as the necessary
input into the formula (\ref{cdilog}). We found that $\C_1$ runs from the
origin to a positive real endpoint $x_1$, with no winding about
$x=-1$, and that $\C_2$ runs from the origin to a point $x_2<-1$,
passing above $x=-1$. The relevant stationary solution to
the Y-system is therefore the $s{=}5$ case
of (\ref{Ysystwosoln}):
\[
x_1=1/y^{(5)}_1\quad,\qquad x_2=1/y^{(5)}_2~.
\]
{}From the contours follow the relevant branches of the logarithms:
first we deduce
\bea
\logci(1+x_1)&=&\log|1+x_1|\,;\nn\\
\logcii(1+x_2)&=&\log|1+x_2| +\imath\pi\,,
\label{logipe}
\eea
and then feeding this into the imaginary part of (\ref{kinktbalim})
establishes that
\bea
\logci (x_1)&=&\log|x_1| -2\imath\pi\,;\nn\\
\logcii (x_2)&=&\log|x_2| -3\imath\pi\,.
\label{logxbe}
\eea
The branch of $\logci(x_1)$ means that, despite $x_1$ being a positive real
number, $L_+(\C_1)$ has an imaginary part. Extracting this and dealing
with the rest via the change of variables $t=x/(1{+}x)$,
\[
L_+(\C_1)=\rdilog(x_1/(1{+}x_1))+\imath\pi\log(1{+}x_1)~.
\]
For $L_+(\C_2)$ the calculation is just as described in section
3.1, modulo the revised branch of $\logcii(x)$. The result is
\[
L_+(\C_2)=-\rdilog(1)-\rdilog(1+1/x_2)-\frac{3}{2}\pi^2
+\frac{1}{2}\imath\pi\log\left[x_2(1{+}x_2)^3\right]~.
\]
Adding the two together, the stationary Y-system can be used to show
that the imaginary part of the total vanishes, and via the sum rule
(\ref{rulethree}) we obtain
\[
c(0)=-68/7
\]
which is indeed the expected answer for the third sheet, given that at
$r{=}0$ it sees the state generated by $\varphi_{11}$.

During the continuation around $\tilde B$, the singularity 
at $\beta^{(0)}_2$ remains in the left-hand kink system, but those at
$\beta^{(0)}_1$ and $\beta^{(1)}_2$ defect to the right-hand system.
In the notation of (\ref{fullc}), the amounts to
\bea
\nu^{(0)}_1~~~=~-1\,,&&\quad \nu^{(0)}_2~~~=~+1\,,\nn\\
\nu^{(j{>}0)}_1=~+1\,,&&\quad \nu^{(1)}_2~~~=~-1\,,\nn\\
&&\quad \nu^{(j{>}1)}_2=~+1\,.
\eea
Although the singularity at $\beta^{(0)}_2$ doesn't change its
allegiance, it does move in a significant way, and in the new kink
limit it comes to be symmetrically placed with the defecting
$-\beta^{(1)}_2$. The relevant data about the asymptotic disposition
of singularities is conveniently summarised using the constants 
$c^{(j)}_a$ defined by (\ref{fullc}):
\bea
\im c^{(0)}_1~~~=~~~0~~\,,&&\quad 
\im c^{(0)}_2~~~=+1.10933(9)\frac{\pi}{5}\,,\nn\\
\im c^{(j{>}0)}_1=\pm\frac{7\pi}{10}\,,&&\quad 
\im c^{(1)}_2~~~=-1.10933(9)\frac{\pi}{5}\,,\nn\\
&&\quad \im c^{(j{>}1)}_2=\pm\frac{7\pi}{10}\,.
\label{singposns}
\eea
In addition
\eq
\re c^{(0)}_2=\re c^{(1)}_2\,.
\en
The imaginary parts of the singularity positions are important
because, via equations (\ref{zlines}) and (\ref{ylines}), they
determine
the directions along which singular lines leave the point $r{=}0$ on 
the Riemann surface. The pattern here on the third sheet is much richer
than previously seen: in the sector $\pi/2>\arg(r)\ge 0$, there are three
type~I lines and two type~II lines. Their directions are:
\bea
\arg(r)= 2.10933(9)\frac{\pi}{5} && \hbox{(\,for $z_2(\theta)\,$);}
\label{dir1}\\
\arg(r)= 1.10933(9)\frac{\pi}{5} && \hbox{(\,for $Y_2(\theta)\,$);}
\label{dir2}\\
\arg(r)= ~~\frac{\pi}{5}~~~~~\,\qquad&& \hbox{(\,for $z_1(\theta)\,$);}
\label{dir3}\\
\arg(r)= 0.10933(9)\frac{\pi}{5} && \hbox{(\,for $z_2(\theta)\,$);}
\label{dir4}\\
\arg(r)= ~~\,0~~~~~~\,\qquad&& \hbox{(\,for $Y_1(\theta)\,$).}
\label{dir5}
\eea
(Just beyond this sector there is, as always, an infinite
accumulation of type~I lines, with
$\arg(r)=7\pi/10-\pi/5=\pi/2$.)
As $r$ moves towards the real axis keeping near to the origin, 
with $\arg(r)$ decreasing from just below $\pi/2$ to zero, the singular
lines are crossed in turn, and various generalised TBA
equations are encountered. First, the basic TBA applies; then, as
the type~I line
(\ref{dir1}) is passed, $\beta^{(1)}_2$ enters the strip
$|\im\beta|<\pi/5$.
The zeroes of $z_2(\theta)$ at
$\pm\theta^{(1)}_2$ cross the real $\theta$ axis and we move into
territory described by equations (\ref{oneTBA}),
(\ref{oneTBAc}) and
(\ref{oneTBAt}). However this time it is the singularities at
$\pm\theta^{(1)}_2$  that have become active, rather than those at
$\pm\theta^{(0)}_2$, and so strictly speaking the labelling in the
generalised TBA equations should be adjusted accordingly. 
This has an interesting consequence. The same region of the Riemann
surface, namely that part of the third sheet described by equations
(\ref{oneTBA}), (\ref{oneTBAc}) and (\ref{oneTBAt}),
can alternatively be reached from
the basic TBA by first continuing around the point $A$, and then
looping anticlockwise around the point $B$. As $A$ is encircled, the
singularities at $\pm\theta^{(0)}_2$ become active, and these remain
the only active singularities during the rest of the
continuation. Their ultimate
positions are those at which the singularities at $\pm\theta^{(1)}_2$
had previously been found,
when the continuation had been about $\tilde B$.  Therefore, if the
first continuation is combined with the reverse of the second, the net
effect is to swap over $\theta^{(0)}_2$ and $\theta^{(1)}_2$. 
In the last subsection it was observed that a singularity can be
swapped with its negative under continuation around certain points on
the Riemann surface; here is an example of a more complicated exchange,
associated with a continuation around a non-trivial cycle on the
surface.

Passing the type~II line (\ref{dir2}), the next singularities, zeroes of
$Y_2(\theta)$ at $\pm\beta^{(1)}_2$, cross the real axis. 
This does not change the set of $\beta^{(j)}_a$ inside the strip 
$|\im\beta|<\pi/5$, and the only effect of the transition is
to reallocate the
active singularities. Equations (\ref{oneTBA}), (\ref{oneTBAc})
and (\ref{oneTBAt}) continue to hold, but now with the active
singularities at $\pm\theta^{(1')}_2(r)=\pm(\theta^{(1)}_2(r){-}2\pi/5)$.

A genuinely new equation is only found when the type~I line
(\ref{dir3}) is crossed. The singularities in $L_1(\theta)$ at
$\pm\theta^{(0)}_1$ become active, and a generalised TBA
equation involving four active singularities results.
It should by now be clear that the labelling of singularities is
path-dependent, and so for notational
simplicity we shall opt to label
the two pairs of active singularities as $\pm\theta^{(0)}_1$ and
$\pm\theta^{(0)}_2$, with $\theta^{(0)}_1$ and $\theta^{(0)}_2$ both
having positive imaginary parts. This means that $\theta^{(0)}_1$
belongs to the right-hand kink system, and $\theta^{(0)}_2$ to the
left-hand one. The new equation is
then
\eq
\ep_a(\theta)=m_a r\cosh\theta
+\logc{S_{a1}(\theta{-}\theta_1^{(0)})\over S_{a1}(\theta{+}\theta_1^{(0)})}
     {S_{a2}(\theta{-}\theta_2^{(0)})\over S_{a2}(\theta{+}\theta_2^{(0)})}
-\sum_{b=1}^{2} \phi_{ab}{*}L_b(\theta)~.
\label{R11}  
\en
with
\eq
c(r)= 
{12r\over\pi}\imath\,(m_1\sinh\theta_1^{(0)}+m_2\sinh\theta_2^{(0)})
 +{3 \over \pi^2} \sum_{a=1}^{2} \iintd\,
 m_a r\cosh\theta L_a(\theta)~.~~
\label{r12}
\en
There are now two independent singularity positions to be fixed, and
the equations for these are found by setting $\theta$ equal to
$\theta^{(0)}_1$ and then $\theta^{(0)}_2$ in (\ref{R11}).
This gives
\eq
 \imath\pi =m_a r\cosh\theta_a^{(0)}
+ \sum_{b=1}^{2} \logc{S_{ab}(\theta_a^{(0)}{-}\theta_b^{(0)})\over 
S_{ab}(\theta_a^{(0)}{+}\theta_b^{(0)})}
-\sum_{b=1}^{2} \phi_{ab}{*}L_b(\theta_a^{(0)})~,~~a=1,2\,.
\label{t2q}  
\en
There are $2n\pi\imath$ ambiguities on the left-hand sides of these
equations, both from the branch choices involved in solving
$\exp(-\ep_a(\theta^{(0)}_a))=-1$, and from the memory of the
particular path of continuation implicit in the notation $\logc$. As
seen in the last subsection, it is important to sort these out if a
direct analytic treatment of the ultraviolet
asymptotics is to be given. We will leave
further discussion of this point for now, contenting ourselves
with the calculation that was performed earlier, in the sector of the third
sheet where the zero-singularity TBA applied. 
These worries are unimportant for numerical work, and using
the exponentiation and inversion idea described earlier, equations
(\ref{R11}) -- (\ref{t2q}) can be solved
over large regions of the complex $r$ plane. In particular,
they turn out 
to hold even when $r$ is real, so long as the infrared regime has
been reached. However near to $r{=}0$ there is one more transition
to go, caused by the type~I line (\ref{dir4}). As this singular
line is passed, the zero of $Y_2(\theta)$ at $\beta^{(0)}_2$
leaves the strip $|\im\beta|<\pi/5$
and the associated singularity ceases to be active. There is now only
one pair of active singularities, and in contrast to the
previous example, these singularities are in $L_1$ and not $L_2$. The
relevant equations 
can be found from (\ref{R11}) -- (\ref{t2q}) by
deleting all terms involving $\theta^{(0)}_2$:
\eq
\ep_a(\theta)=m_a r\cosh\theta
+\logc{S_{a1}(\theta{-}\theta_1^{(0)})\over S_{a1}(\theta{+}\theta_1^{(0)})}
-\sum_{b=1}^{2} \phi_{ab}{*}L_b(\theta)~,
\label{R111}  
\en
\eq
c(r)= 
{12r\over\pi}\imath\,\sinh\theta_1^{(0)}
 +{3 \over \pi^2} \sum_{a=1}^{2} \iintd\,
 m_a r\cosh\theta L_a(\theta)~,~~
\label{r121}
\en
and
\eq
 \imath\pi =m_1 r\cosh\theta_1^{(0)}
+\logc{S_{11}(0)\over 
S_{11}(2\theta_1^{(0)})}
-\sum_{b=1}^{2} \phi_{ab}{*}L_b(\theta_a^{(0)})~.~~
\label{t2q1}  
\en
These equations should hold in the 
sector $0.10933(9)\pi/5>\arg(r)>0$ while $|r|$ remains small enough that
the splitting of the pseudoenergies into left and right kink systems holds.
However the proximity of
the type~II singular line (\ref{dir5}), and the resulting presence of
zeroes of $Y_1(\theta)$ near
to the real $\theta$ axis, means
that a direct numerical treatment fails, at least with the methods
that we have at present. Fortunately 
the equations can be recast into a more tractable form.
As before, the key is to use the identity (\ref{lSresult}) to
eliminate some of the singular behaviour in $\ep_a$ and
$L_a$. Here, the $d{=}1$ case of (\ref{lSresult}) is relevant, and
the revised definitions of $\ep_a$ and $L_a$ are
\eq
\hat\ep_a(\theta)= \ep_a(\theta) -
l^{[T_2]}_{a1}
\logc\left(\sigma_h(\theta{-}\beta_1^{(0)})
\sigma_h(\theta{+}\beta_1^{(0)})\right)
\label{modaa}
\en
and 
\eq
\hat L_a(\theta)= \logc\lf(\lf(
\sigma_h(\theta{-}\beta_1^{(0)}) 
\sigma_h(\theta{+}\beta_1^{(0)})\ri)^{l^{[T_2]}_{a1}}
+ e^{-\hat\ep_a(\theta)} \ri)\,.
\label{modbb}
\en
Only $\ep_2$ and $L_2$ are changed by this manoeuvre, as
expected given (\ref{y2tab}). The generalised TBA then falls into the
more traditional form of equations (\ref{oneTBAg}) and
(\ref{oneTBAgc}), with of course the proviso that the singularities
must be correctly placed. This equation turns out to be valid
for real $r$ out to $r=r_{c_2}\approx 4.7271(0)$; some numerical 
results are compared with TCSA data in table~\ref{tabtwouv} of
appendix~\ref{numapp}.

\begin{figure}[tbh]
\epsfxsize=11cm
\quad\epsfbox{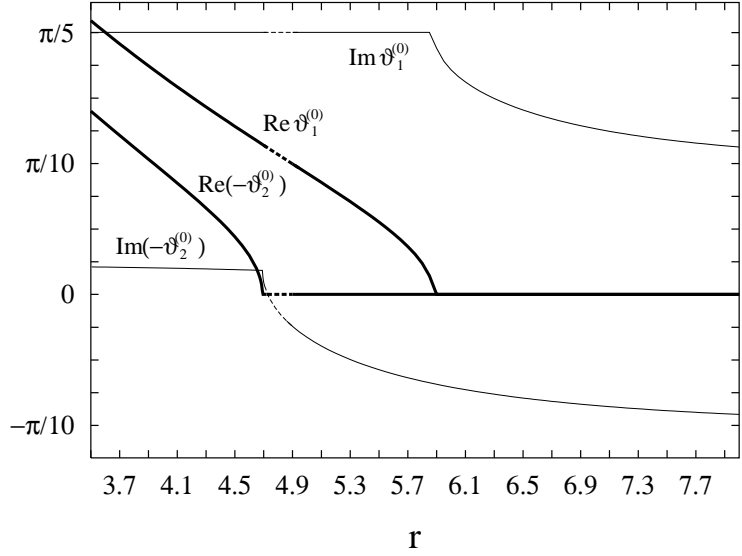}
\vskip -10pt
\caption{\leftskip=.5cm  {\protect \small Singularity movement for
the second one-particle state: the dotted segments indicate the region
where the numerics failed to converge, and were obtained by
extrapolation from larger values of $r$.
Transitions are at
$r_{c_1}= 4.691(0)$, $r_{c_2}= 4.7271(0)$, and $r_{c_3}= 5.8786(1)$.
}
\rightskip=0.5cm }  
\label{singplot}
\end{figure}

To understand why the equation breaks down at $r=r_{c_2}$, we tracked
the motion of the first few singularities as $r$ varied on the segment
$[0,r_{c_2}]$. At $r=0$, their positions follow from the kink system.
Labelling the singularities consistently with (\ref{R11}), the
right-hand kink system contains amongst other things
an active singularity at
$\theta^{(0)}_1$, with imaginary part equal to $\imath\pi/5$, and
a pair of
conjugately-placed inactive singularities at $-\theta^{(0)}_2$ and
$\theta^{(1)}_2$, with imaginary parts equal to $0.10933(9)\pi/5$ and
$-0.10933(9)\pi/5$ respectively. As $r$ increases, the real part of
$-\theta^{(0)}_2$ decreases, until at $r=r_{c_1}=4.691(0)$ it
vanishes completely and the singularity at $-\theta^{(0)}_2$, part of
the right-hand kink system, collides with one at $-\theta^{(1)}_2$,
coming in from the left-hand system. At this stage the imaginary parts
of these singularities are still positive. During the collision, the
singularities
undergo a 90 degree `scattering' and for $r$ just larger than
$r_{c_1}$, they have turned into a pair of 
singularities on the positive imaginary axis,
one heading up, and one heading down towards $\theta{=}0$. Choosing
to allocate the label $-\theta^{(0)}_2$ to the second of these, we
were able to track its progress down the imaginary $\theta$ axis for a
short way before our numerical methods became unstable. Extrapolating beyond
this point and also making use of data from larger values of $r$, we
found that $-\theta^{(0)}_2$ hits the origin when $r$ arrives
at $r_{c_2}$.
The natural explanation for this behaviour is that the
type~I line (\ref{dir4}), which started at
the origin moving away from the real $r$ axis, has curved round and
hit the real axis again.  (It then returns to the origin
on the complex-conjugated path in the lower half plane.)  We confirmed
this picture by tracking singularity positions for complex values of
$r$, and conclude that the domain of applicability of equations 
(\ref{R111}) -- (\ref{t2q1}) is restricted to
a small neighbourhood of the segment $[0,r_{c_2}]$ of the real axis. 
As this region is left, the singularities at $\pm\theta^{(0)}_2$ cross
the real $\theta$ axis and the more complicated equations (\ref{R11}) --
(\ref{t2q1}) come into operation. 
The singularity movement described so far is plotted as the lines
labelled $\re(-\theta^{(0)}_2)$ and
$\im(-\theta^{(0)}_2)$ on figure~\ref{singplot}, showing first the
vanishing of the real part at $r_{c_1}$, and then 
the crossing of $-\theta^{(0)}_2$ into the lower half plane at $r_{c_2}$.
\setlength{\unitlength}{1.mm}
Looking elsewhere on the figure, we see that $\im\theta^{(0)}_1$
remains equal to $\pi/5$ as the point $r_{c_2}$ is passed. This means
that the singularities at $\pm\beta^{(0)}_1$ are still on the real $\theta$
axis, and the transition to the infrared regime is not quite complete.
For numerical work, we must continue to use the modified definitions
(\ref{modaa}) and (\ref{modbb}) 
to cope with these singularities.
Whilst it is only the singularities at
$\pm\beta^{(0)}_1$ that have to be dealt with in this way, 
there are now also active
singularities at $\pm\theta^{(0)}_2$, and
the equations are more
elegant if the associated
points $\pm\beta^{(0)}_2$ are `desingularised' at the
same time: setting
\eq
\hat\ep_a(\theta)= \ep_a(\theta) -
\logc\prod_{d=1}^2\left(\sigma(\theta{-}\beta_d^{(0)})
\sigma(\theta{+}\beta_d^{(0)})\right)^{l_{ad}^{[T_2]}}
\label{twomedTBAdefa}
\en
and
\eq
\hat{L}_a(\theta)= \logc\lf( 
\prod_{d=1}^2\left(\sigma(\theta{-}\beta_d^{(0)})
\sigma(\theta{+}\beta_d^{(0)})\right)^{l_{ad}^{[T_2]}}
+ e^{-\hat{\ep}_a(\theta)} \ri)\,,
\label{twomedTBAdefb}
\en
they again acquire the form of equations (\ref{oneTBAg}) and
(\ref{oneTBAgc}). This time, though, there are two independent singularity
positions to fix:
\bea
e^{\hat\ep_1(\theta^{(0)}_1)}&=&
-\imath\coth\lf(\fract{5}{2}\theta^{(0)}_1+\fract{\imath\pi}{4}\ri)\,;\nn\\
e^{\hat\ep_2(\theta^{(0)}_2)}&=&
-\imath\coth\lf(\fract{5}{2}\theta^{(0)}_2+\fract{\imath\pi}{4}\ri)\,.
\label{twoTBAgt}
\eea
These equations were used to obtain 
tables \ref{tabtwomed} and \ref{tabtwoir} of appendix~\ref{numapp};
the second set was also confirmed by the direct use of 
the system (\ref{R11}) -- (\ref{t2q}). As can be seen from
figure \ref{singplot}, there is a
final transition to the infrared regime, at
$r=r_{c_3}\approx 5.8786(1)$, where the real part of $\theta_1^{(0)}$
vanishes and all active singularities take up positions on the
imaginary axis. 

As $r$ grows further, both $\theta^{(0)}_1$ and $\theta^{(0)}_2$
approach $\imath\pi/10$, and the identity
\eq
m_2=2\sin(\frac{\pi}{10})(m_1{+}m_2)
\label{mass2}
\en
leads to the result $c(r)\sim -6m_2r/\pi$, or $E(\lambda,R)\sim E_{\rm
bulk}(\lambda,R)+M_2$. This is as expected, and just as before we can
supplement the result by calculating two further corrections. For
$a=1,2$ we set
$\theta^{(0)}_a(r)=\imath\pi/10+\delta_a(r)$, where both $\delta_1(r)$
and $\delta_2(r)$
are purely imaginary for real $r$ larger than $r_{c_3}$, and tend to
zero as $r\rightarrow\infty$. Imposing (\ref{t2q}), with the
convolution terms ignored as being subleading, leads to the following
pair of conditions:
\[
 {S_{12}(\delta_1{-}\delta_2) \over 
S_{11}(\imath \pi/5{+}2\delta_1)S_{12}(\imath\pi/5{+}\delta_1{+}\delta_2)}
\sim  e^{-m_1r\cos(\pi/10)}~;
\]
\[
{S_{12}(\delta_2{-}\delta_1) \over 
S_{12}(\imath \pi/5{+}\delta_1{+}\delta_2)S_{22}(\imath\pi/5{+}2\delta_2)} 
\sim e^{-m_2r\cos(\pi/10)}~.
\]
(Note that with the equations exponentiated in this way the branch
ambiguities have gone.) It will be convenient to have a notation for
the most singular parts of the
various poles controlling these equations:
given a function $f(\theta)$,
we let $\R f(\theta_0)$ denote the leading coefficient of its Laurent
expansion about $\theta_0$:
\eq
f(\theta)=\R f(\theta_0)(\theta{-}\theta_0)^p+
\sum_{k=p{+}1}^{\infty}A_k(\theta{-}\theta_0)^k~;\quad 
\R f(\theta_0)\neq 0\,.
\label{laurcoeff}
\en
(If the pole at $\theta_0$ is simple, then
$\R f(\theta_0)$ is just the residue.)
The limiting behaviour of $\delta_1$ and $\delta_2$ is then determined
by
\bea
\frac{1}{S_{11}(\imath\pi/5)}\,
\frac{\delta_1{+}\delta_2}{\R S_{12}(\imath\pi/5)}
&\sim&  e^{-m_1r\cos(\pi/10)}~; \nn\\
\frac{\delta_1{+}\delta_2}
{\R S_{12}(\imath\pi/5)}\,
\frac{2\delta_2}{\R S_{22}(\imath\pi/5)} 
&\sim& e^{-m_2r\cos(\pi/10)}~.\nn
\eea
Solving, and using (\ref{sdef}) for the S-matrix elements,
we have
\bea
\delta_2(r)&\sim&
-\imath\tan(\frac{2\pi}{5})\,e^{-(m_2{-}m_1)r\cos(\pi/10)},\nn\\
\delta_1(r)&\sim&
\,2\imath\tan(\frac{3\pi}{10})\tan(\frac{2\pi}{5})^2\,
e^{-m_1r\cos(\pi/10)}-\delta_2(r)\,. \nn
\eea
Note that $\imath\pi/10$ is approached
from above by $\theta^{(0)}$, and from below by
$\theta^{(0)}_2$. The leading effect that this has on
the value of $c(r)$ is through the term $\frac{12
r}{\pi}\imath(m_1\sinh\theta^{(0)}_1{+}m_2\sinh\theta^{(0)}_2)$ 
in (\ref{r12}). The result matches the contributions of the 
field-theoretic $\mu$-terms, given in appendix~\ref{ftres}. Notice that
this time there are two distinct exponentials, reflecting the two
processes $2\rightarrow 1\,2\rightarrow 2$ and $2\rightarrow
1\,1\rightarrow 2$.
A different type of correction comes from the thus-far
neglected integral in (\ref{r12}), using the first two
terms on the right-hand side of (\ref{R11}) to estimate
$L_a(\theta)$.  The modifications
to $\theta^{(0)}_1$ and $\theta^{(0)}_2$ have a subleading effect on
this term, and so their asymptotic 
values of $\imath\pi/10$ can be used, whereupon the following identity
becomes relevant:
\eq
\frac{S_{a1}(\theta{+}\frac{\imath\pi}{10})}
     {S_{a1}(\theta{-}\frac{\imath\pi}{10})}
\frac{S_{a2}(\theta{+}\frac{\imath\pi}{10})}
     {S_{a2}(\theta{-}\frac{\imath\pi}{10})}
= S_{a2}(\theta+\frac{\imath\pi}{2})~.
\label{ss2}
\en
The final result is
\bea
c(r)\!\! &\sim& \!\! { -6 r \over \pi}\Bigl(m_2 
+2C_1m_1e^{-m_1r\sin(2\pi/5)}
+C_2m_1e^{-m_1r\sin(\pi/5)} \nn \\
 &&\qquad\qquad\!\! -{1\over 2\pi}\sum_{a=1}^{2}\iintd\,
 m_a \cosh\theta~ S_{2a}(\theta+{\imath \pi\over 2})e^{-rm_a\!\cosh 
\theta }\Bigr)~.\qquad~~~
\eea
where
\eq
C_1=2\cos(\frac{\pi}{10})\tan(\frac{3\pi}{10})\tan(\frac{2\pi}{5})^2
 =24.79837\dots
\en
and
\eq
C_2=-2\sin(\frac{\pi}{5})\tan(\frac{2\pi}{5})
=-3.61803\dots
\en
and the identity
\eq
(m_2-m_1)\cos(\pi/10)=m_1\sin(\pi/5)
\en
was used to rewrite the second $\mu$-term. These match the
predictions from field theory recorded in table~2 of \cite{KMd}, or
appendix~\ref{ftres}.

%
\resection{Generalisations}
%
In this section we indicate how some of the above results for the $T_2$
model can be extended to the rest of the $T_N$ series. Most of our
analysis is confined to the infrared region, where an elegant
pattern emerges. Even for real $r$, the story becomes more
complicated outside this region, as
the movement of singularities causes the
equations to change their forms.
The simplest examples were seen in refs.~\cite{BLZa,DTa} and also
in section \ref{oneTBAsec} above, with the single transition that could be
traced to the arrival of a pair of singularities exactly on the real $\theta$
axis. As was seen in that section,  the effect is rather mild -- in
particular, the infrared equations continue to apply if $r$ is given
an infinitesimal imaginary part. 
The case studied in section \ref{thirdsheetsec}, the
second 1-particle state of the $T_2$ theory, provides the first
example of the more complicated behaviour that we expect to see in the
general case:
one pair of active singularities, present in the infrared, disappears
completely in the ultraviolet.
For general $N$ we will see that the inverse  process is also possible: a 
certain number of new singularities can become active
at small $r$. This will be described at the end of
the section, using an example from the $T_4$ theory.

Our initial  conjecture is based on the behaviour
already observed for $N{=}1$ and $N{=}2$, and the following two
identities. If we set
\[
B=N-A+1
\]
then
\eq
2 \sin {\pi \over 2 h} \sum_{b=B}^{N}\! m_b = m_A
\label{massfor}
\en
and
\eq
\prod_{b=B}^{N} 
\frac{S_{ab}(\theta + \imath {\pi \over 2 h})}{S_{ab}(\theta
 - \imath {\pi \over 2 h})} 
=S_{Aa}(\theta+ \imath { \pi \over 2})
\label{sfor}
\en
These generalise equations (\ref{mass1}),(\ref{ss1}) and
(\ref{mass2}),(\ref{ss2}) respectively,
and make natural the supposition
that when $r$ is sufficiently large, the 
$A^{\rm th}$ one-particle state is exactly described by a generalised TBA
equation with $A$ pairs of active singularities, these singularities
being found in the pseudoenergies $\ep_b(\theta)$, $b=B\dots
N$, with all of them tending to $\imath\pi/2h$ in the infrared
limit. Assuming that this singularity pattern can be achieved through
some path of continuation from the ground state,
the generalised TBA equation must then be
\eq
\ep_a(\theta)=m_a r\cosh\theta
+\!\sum_{b=B}^{N}\logc{S_{ab}(\theta{-}\theta_b^{(0)})\over 
S_{ab}(\theta{+}\theta_b^{(0)})}
-\sum_{b=1}^{N} \phi_{ab}{*}L_b(\theta)~,
\label{tne}  
\en
with the active singularity positions $\theta^{(0)}_a$,
$a=B\dots N$, determined by
the equations
\eq
 \imath \pi =m_a r\cosh\theta_a^{(0)}
+ \sum_{b=B}^{N} \logc{S_{ab}(\theta_a^{(0)}{-}\theta_b^{(0)})\over 
S_{ab}(\theta_a^{(0)}{+}\theta_b^{(0)})}
-\sum_{b=1}^{N} \phi_{ab}{*}L_b(\theta_a^{(0)})~,
\label{tnq}  
\en
and $c(r)$ given by the formula
\eq
c(r)=  
{12r\over\pi}\imath\,(\sum_{b=B}^{N}\!m_b\sinh\theta_b^{(0)})
 +{3 \over \pi^2} \sum_{a=1}^{N} \iintd\,
 m_a r\cosh\theta L_a(\theta)~.~~
\label{tnc}
\en
At large $r$ the dominant term is found by setting
$\theta_a^{(0)}= \imath \pi/2h$ and using 
(\ref{massfor}), giving  $c(r) \sim -6 m_{A} r/\pi$ or $E(\lambda,R) 
\sim E_{\rm bulk}(\lambda,R) +M_{A}\,$. 
This is correct for a one-particle state 
with mass $m_A$. A couple of further corrections can be obtained
exactly.
The simplest to find is the $F$-term, and this
comes on using the asymptotic 
$\ep_a(\theta) \sim m_a r \cosh \theta - \logc S_{aA}(\theta{+}\imath 
\pi/2)$ in  (\ref{tnc}) (equation (\ref{sfor}) having been used 
in order to simplify the ratio of S-matrices appearing in (\ref{tne})~).  
This gives  the following 
correction to $c(\infty)$:
\eq 
{3r\over\pi^2}\sum_{b=1}^N\iintd\,
m_b\cosh\theta~S_{Ab}(\theta+{\imath\pi\over 2})e^{-m_br\cosh\theta}.
\label{scorr}
\en
With the leading asymptotic of the ground state subtracted off, this
matches the $F$-term (\ref{delta1}).
The derivation of the second correction is more tedious. 
We set $\theta^{(0)}_a=\imath\pi/2 h+\delta_a$, with $\delta_a$ purely
imaginary, and
start from an infrared form of equation (\ref{tnq}), obtained by 
dropping the convolution term and then exponentiating:
\eq
\prod_{b=B}^N\frac{S_{ab}(\delta_a{-}\delta_b)}
{S_{ab}(\imath\pi/h{+}\delta_a{+}\delta_b)}
\sim -e^{- r m_a \cosh\theta_{a}^{(0)}}\,,\qquad 
a=B,B{+}1,\dots N\,.
\label{expp}
\en
A simple consideration of the signs of the terms in these equations, using
\[
\sign S_{ab}(\imath x)=
\left\{ \begin{array}{ll}
(-1)^{\delta_{ab}}&\quad-\pi/h<x<\pi/h \\
(-1)^{\delta_{ab}+l_{ab}^{[T_N]}}&\quad~~\pi/h<x<2\pi/h
\end{array} \right.
\]
for $x$ real,
and the explicit form of $l^{[T_N]}_{ab}$, establishes that
\[\qquad
\sign\frac{\delta_{a}}{\imath}=(-1)^{B+a},\quad\qquad a=B,B{+}1,\dots N\,,
\]
and
\[
|\delta_{B}| > |\delta_{B+1}|>\dots>|\delta_N|\,.
\]
However more precision is needed if we are to capture the asymptotic.
Noting that 
only the S-matrix elements with $l_{ab}^{[T_N]} \ne 0$ are 
singular at $\theta=\imath\pi/h$, the dominant behaviour of (\ref{expp})
is
\[
\prod_{b=B}^N (\delta_a+\delta_b)^{l_{ab}^{[T_N]}} \sim
-e^{-m_ar\cos(\pi/2h)} 
\R\lf(\prod_{b=B}^N  
\frac{S_{ab}(\imath{\pi\over h})}{S_{ab}(0)}\ri)  \,,
\]
with $\R$ as defined in equation (\ref{laurcoeff}).
Now (\ref{sfor}) can be used at $\theta=\imath\pi/2h$ to simplify the
right-hand side, giving
\[
\prod_{b=B}^N
(\delta_a+\delta_b)^{l_{ab}^{[T_N]}}\sim -e^{-m_ar\cos(\pi/2h)}   
\R S_{Aa}(\imath\fract{h+1}{2h}\pi)\,,\quad a=B,B{+}1,\dots N\,.
\]
Explicitly this reads
\bea
\openup 1\jot
(\delta_B+\delta_{B+1})&\sim&~{-e}^{-m_Br\cos(\pi/2h)}~\R
S_{AB}(\imath\fract{h+1}{2h}\pi) \nn\\
(\delta_{B+1}+\delta_B)(\delta_{B+1}+\delta_{B+2})
&\sim&{-e}^{-m_{B+1}r\cos(\pi/2h)}\R
S_{A,B+1}(\imath\fract{h+1}{2h}\pi) \nn\\
\vdots\qquad\qquad\quad\vdots\qquad&&\qquad~~\vdots\qquad\qquad\qquad\vdots
 \nn\\
(\delta_{N-1}+\delta_{N-2})(\delta_{N-1}+\delta_N)
&\sim&{-e}^{-m_{N-1}r\cos(\pi/2h)}\R
S_{A,N-1}(\imath\fract{h+1}{2h}\pi) \nn\\
(\delta_N+\delta_{N-1})(2\delta_N)
&\sim&~{-e}^{-m_Nr\cos(\pi/2h)}~\R
S_{A,N}(\imath\fract{h+1}{2h}\pi) \nn
\eea
Now a chain of substitutions gives
\bea
\openup 1\jot
{~}\qquad\qquad\quad
\delta_b+\delta_{b+1}&\sim&~~K_be^{-\mu_br}\,,\qquad\qquad b=B,B{+}1,
\dots N{-}1\,;\nn\\
\delta_N&\sim&\halft K_Ne^{-\mu_Nr}\,,\nn
\eea
with
\[
\mu_b = \cos(\pi/2h)\sum_{c=B}^b(-1)^{b+c}m_c
\]
and
\[
K_b=\R\lf(\prod_{c=B}^b\lf(-S_{Ac}(\imath\fract{h+1}{2h}
\pi)\ri)^{(-1)^{b+c}}\ri)\,.
\]
These values will modify $c(r)$, at leading order, through a correction
term equal to
\bea
\openup 1\jot
\Delta_{\mu}c(r)
&=&\frac{12r}{\pi}\imath\sum_{b=B}^Nm_b\cos(\pi/2h)\delta_b\nn\\
&=&\frac{12r}{\pi}\imath
\lf[\sum_{b=B}^{N-1}\mu_b(\delta_b{+}\delta_{b+1})+\mu_N\delta_N\ri]\nn\\
&=&\frac{6r}{\pi}\imath
\lf[\sum_{b=B}^{N-1}2\mu_bK_be^{-\mu_br}+\mu_NK_Ne^{-\mu_Nr}\ri]\,.
\label{mucalc}
\eea
The calculation now splits into two. First, consider a term in the final
sum for which
$b{-}B$ is odd, equal to $2k{-}1$. Then we can use the following mass
relation:
\[
\cos(\pi/2h)\sum_{c=B}^{B+2k-1}(-1)^{B+c+1}m_c=m_k\sin(\fract{A{-}k}{h}\pi)\,,
\]
or
\[
\mu_{B+2k-1}=\mu_{A,k,A-k}\,,
\]
where the notation $\mu_{A,k,A-k}$ is as used in equations (\ref{delta2})
and (\ref{muone}).
Recalling that $\Delta
E(R)=-\frac{\pi}{6R}\Delta c(r)=-\frac{M_1\pi}{6r}\Delta c(r)$
and comparing with the formulae of appendix~\ref{ftres},
we see that the correction due to the $\alpha$-process 
$A\rightarrow k,A{-}k\rightarrow A$ will be reproduced, 
so long as the following
identity between S-matrix residues holds:
\[
-\R\lf(\prod_{c=B}^{B+2k-1}\lf(S_{Ac}
\lf(\imath\fract{h+1}{2h} \pi\ri) 
\ri)^{(-1)^{B+c+1}}\ri) =
\R S_{Ak}(\imath\fract{A{-}k}{h}\pi)\,.
\]
We checked this algebraically up to $N{=}5$, and then numerically to
$N{=}15$. Up to a sign ambiguity encountered when taking a square root, 
the identity can also be deduced
from the following curious property of the S-matrix elements:
\[
\prod_{c=B}^{B+2k-1}   
\lf(S_{Ac}(\theta+\imath\fract{h+1}{2h}\pi)
S_{Ac}(\theta+\imath\fract{h-1}{2h}\pi)\ri)^{(-1)^{B+c+1}} 
=
{}~\frac{S_{Ak}(\theta+\imath\fract{{A-k}}{h}\pi)}{
S_{Ak}(\theta-\imath\fract{{A-k}}{h}\pi)}~.
\]

If $b{-}B=2k$ is even, the story is much the same. The relevant mass
relation is
\[
\cos(\pi/2h)\sum_{c=B}^{B+2k}(-1)^{B+c}m_c=
m_{N-k}\sin(\fract{A{+}N{-}k}{h}\pi)\,,
\]
or
\[
\mu_{B+2k}=\mu_{A,N-k,N+1-A+k}\,.
\]
This has a chance to match the $\beta$ correction (\ref{mutwo}),
from the process $A\rightarrow N{-}k,N{+}1{-}A{+}k\rightarrow
A$. The necessary residue identity is
\[
\R\lf(\prod_{c=B}^{B+2k}\lf(S_{Ac}
\lf(\imath\fract{h+1}{2h} \pi\ri) 
\ri)^{(-1)^{B+c}}\ri) =
\R S_{A,N-k}(\imath\fract{A{+}N{-}k}{h}\pi)\,,
\]
which follows, again up to a sign, from another property of the S-matrix
elements:
\[
\prod_{c=B}^{B+2k}   
\lf(S_{Ac}(\theta+\imath\fract{h+1}{2h}\pi)
S_{Ac}(\theta+\imath\fract{h-1}{2h}\pi)\ri)^{(-1)^{B+c}} 
=
{}~\frac{S_{Ak}(\theta+\imath\fract{{A+N-k}}{h}\pi)}{
S_{Ak}(\theta-\imath\fract{{A+N-k}}{h}\pi)}~.
\]

To finish, note that the term in (\ref{mucalc}) proportional to $K_N$
receives special treatment, in that it is without the factor
of $2$ that multiplies all of the others. If $A{=}2k$, then $b{-}B$ is 
odd for this term, and it corresponds to the $\alpha$-process
$A\rightarrow k,k\rightarrow A$, whilst if $A=2k{+}1$, $b{-}B$ is even and
the term matches the $\beta$-process $A\rightarrow
N{-}k,N{-}k\rightarrow A$. In both cases the missing $2$ is
accounted for by a symmetry factor.

The intricate mechanism by which all of the $\mu$-terms are successfully
recovered leaves little doubt that equations (\ref{tne}) -- (\ref{tnc}) 
describe  exactly the zero-momentum one-particle states at suitably-large
values of $r$. Transitions are inevitable as the ultraviolet regime is
approached, and we have already seen examples of the
sort of
phenomena that can occur. This section ends with one more, showing yet
another way in which the equations can change their form in the crossover
region.

\begin{figure}[htbp]
\epsfxsize=11cm
\epsfbox{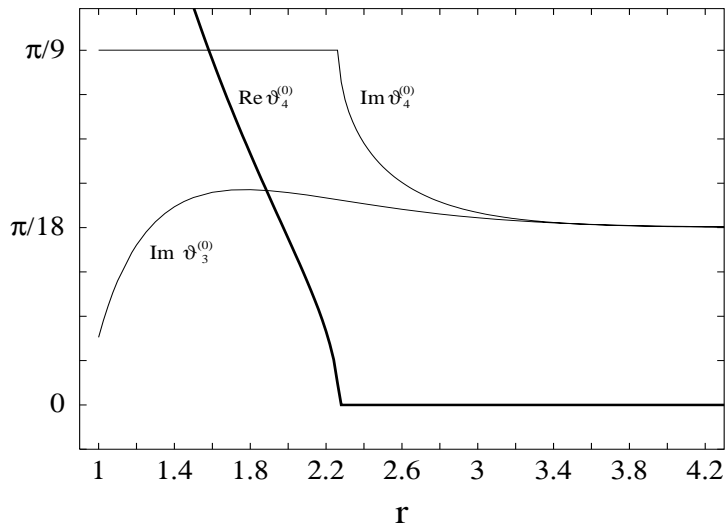}
\caption{\leftskip=.5cm  {\protect \small Singularity movement for
the first one-particle state in $T_4$ ($\re\theta^{(0)}_3$ is equal to
$0$ for all $r$ in the range plotted, and so this line has been omitted).
}
\rightskip=0.5cm }  
\label{singplot1}
\end{figure}

Consider the first one-particle state in the $T_4$ theory. As just
explained, at large $r$ we expect this to be described by a generalised
TBA equation with one pair of active singularities, zeroes of
$z_4(\theta)$. These are at $\pm\theta^{(0)}_4$, with $\theta^{(0)}_4$
purely imaginary, tending to $\imath\pi/18$ from above as $r$ grows.
Conversely, as $r$ decreases our numerical solutions indicate that
$\im\theta^{(0)}_4$ increases towards $\pi/9$, until at 
$r_c\approx 2.266315266(7)$
there is a square root singularity in its value as a function of $r$. 
This transition is shown
in figure~\ref{singplot1}, and is of the simplest kind seen
in previous sections. 
\setlength{\unitlength}{1.mm}
Decreasing $r$ further, $\theta^{(0)}_4$
starts to run  along the line $\im\theta= \pi/9$, with accompanying
zeroes in $Y_3$ and in $Y_4$, both at $\beta^{(0)}_4$, running along the
real $\theta$ axis.
But more importantly, the  numerical solution reveals 
that a so-far inactive singularity also comes into play.
This is a zero of $z_3(\theta)$, at $\theta^{(0)}_3$ say,
which for large $r$ approaches $\imath 
\pi/18$ from above. As $r$ becomes smaller, $\im\theta^{(0)}_3$ at first
grows, reaching a maximum at
$r\approx 1.8$, and then decreases, as shown on figure~\ref{singplot1}.
At around $r=1$, our program became unstable, 
but extrapolation predicts that at $r\approx 0.9234$ the 
singularities at $\pm\theta^{(0)}_3$ will cross the real axis and become
active.
At this stage $\theta^{(0)}_3$ is still on the imaginary $\theta$ axis, 
and a
natural scenario for its subsequent movement is that it ultimately
reaches $-\imath \pi /9$, undergoes a transition there
and finally
runs along the line $\im\theta=-\pi/9$.
The ultraviolet solutions for $r$ real
would then have both $\beta^{(0)}_3$ and $\beta^{(0)}_4$
sitting on the real axis, 
$Y_4(\theta)$ having
four zeroes on the real axis, at $\pm \beta_3^{(0)}$ and $\pm 
\beta_4^{(0)}$, and $Y_3(\theta)$ having two, at
$\pm \beta_4^{(0)}$. Via the Y-system, the zeroes of $Y_3(\theta)$ 
also force 
$Y_2(\theta)$ to vanish at $\pm\beta_3^{(0)}$.
Unfortunately, we have not yet developed the computer code necessary to
verify this picture directly. In its absence we can at least make a couple
of checks. First, the zeroes just described tell us
the signs of the numbers $y_a=Y^{\rm kink}_a(-\infty)$ for the kink versions
of the solution under discussion, since we already know that $Y^{\rm
kink}(+\infty)=+\infty$:
\[
\sign (y_1,y_2,y_3,y_4) =(+1,-1,-1,+1)~.
\]
These match the $s{=}3$, $N{=}4$ case of the solutions
\[
y^{(s)}_a=\frac{\sin(sa\pi/(h{+}2))\,\sin(s(a{+}2)\pi/(h{+}2))}
{\sin(s\pi/(h{+}2))^2}\qquad(\,h=2N{+}1\,)
\]
to the stationary $T_N$ Y-system, which generalise (\ref{Ysystwosoln}).
The second check in some senses subsumes the first. 
Recall from the $T_2$ discussions
that the kink system for the first generalised
TBA equation, studied in section~\ref{oneTBAsec}, was alternatively found
using the second, excited solution to 
the ground-state TBA equation on the
positive-$\lambda$ line. For $T_4$ we found a similar multiplicity of
solutions to the ground-state equation. An excited solution to the basic
$T_4$ TBA equation can be found by continuing once anticlockwise around the
point $r_0=\rho_0e^{11\pi\imath/36}$, $\rho_0\approx 1.665(9)$.
The limiting singularity pattern of this
solution on the positive-$\lambda$ line 
$r=\rho e^{11\pi\imath/36}$
near to $\rho=0$ turns out to
match precisely with the predictions just
made on the basis of the assumed transitions in the generalised equation 
along the real $r$ axis.

Thus not only do singularities become inactive as the ultraviolet is
approached, as seen in section~\ref{thirdsheetsec}, but also previously
inactive singularities can become active.
More work will be needed to unravel the full pattern of
these transitions, but we have at least exhibited examples of the 
full range of possibilities.

%
\resection{Conclusions}
The theories studied in this paper provide a useful set of toy
models, on which it has been possible to test and develop the method of 
analytic continuation of TBA equations advocated in \cite{DTa}.
 
Although relations with
the sine-Gordon  model at $\beta^2/8\pi=2/(2N{+}3)$ \cite{Sa}, and with
the
$Z_{2N{+}1}$-symmetric conformal theories  perturbed by their
first thermal operators~\cite{Zb,KMa}, lend the $T_N$ theories some additional
physical interest in
their own right, the main objective has been to find general
patterns of wider applicability. In the following, we summarise the main
results obtained, and indicate some open questions that remain.
\begin{itemize}
\item
{\bf Shifted conjugation symmetry:} 
The analyticity of the $Y_a$'s
as functions of the  
variables $a_{+}$ and $a_{-}$, near to $a_+=a_-=0$, was conjectured in
section~2, and as a 
consequence the  shifted conjugation symmetries (\ref{selfconj}) and
(\ref{yconj}) were deduced.
These enabled the TBA equations to be studied in regions
where the direct numerical approach failed to converge. 
In particular, we found in this way that the branch points
connecting the various sheets of the Riemann surface are not pinch
singularities, but rather arise from the multivalued nature of solutions to
individual TBA equations. This contrasts with the `free'
behaviour found for the thermally-perturbed Ising model.
\item
{\bf Singular lines: }
Type~I and type~II singular lines were defined in section~\ref{singlsec}.
The type~I lines divide the Riemann surface into disjoint regions, within
each of which a different generalised TBA equation holds sway. The type~II
lines are less significant: the
TBA equations before and after continuation through these lines can be
recast into equivalent forms. 
As explained in section~\ref{typetwosec}, this is reflected in the fact
that the crossing of a type~I line changes the number of zeroes of the
$Y_a(\theta)$ in the strip $|\im\theta|<\pi/h$, whilst the crossing of a
type~II line does not.
Mapping the pattern of singular lines gives
an insight into the relationships between the various generalised TBA
systems, and a number of techniques were explored to help in this task. In
particular, the result described at the end of section~\ref{singlsec},
based on the ultraviolet splitting of pseudoenergies into kink systems,
allows the asymptotic pattern of lines at small $r$ to be controlled on any
sheet, just from knowledge of the singularity locations for
the relevant set of kink pseudoenergies. So far we have only studied this
in detail for the $T_N$ systems, but it is clear that the same picture will
hold for any TBA system which splits into separated kink systems in the
ultraviolet regime.
\item
{\bf Exotic solutions to the basic TBA equation: }
A basic TBA equation, with no active singularities, has direct access to a
series of branch points in the scaling functions. These always turn out to
be approximately lined up along the imaginary $r$ axis, as explained for
the case of the SLYM in section~3 of ref.~\cite{DTa}. (Figure~4 of that paper
shows two of these points for the SLYM, whilst the points $A$, $\tilde B$
and $\tilde D$ on figure~\ref{surface} are three examples from the $T_2$
theory.) Continuing $r$ anticlockwise around these points, we remain in
the domain of validity of the basic TBA, and uncover some previously
unsuspected solutions. Furthermore, by returning $r$ to the origin along
lines of constant argument, we found, in sections~\ref{basicsec}
and~\ref{thirdsheetsec}, the `excited' kink solutions on the second and
third sheets of the $T_2$ Riemann surface, without ever having to solve a
generalised TBA equation. It would be interesting to know whether {\it all}
kink solutions, both in these and other models, can be obtained in this way.
\item
{\bf Analytic continuation and sum rules: }
Continuation in $r$ allows the contours taken in the generalised
dilogarithm functions $L_+(\C_a)$ to be determined unambiguously, and in a way
which should work completely generally. This has the potential to provide a
more `physical' justification for the continuations
discussed as purely mathematical properties of dilogarithmic sum rules
in~\cite{KNa}, and deserves to be studied in greater depth.
\item
{\bf Desingularisation of the generalised equations: }
By explicitly removing the singularities of the pseudoenergies
$\ep_a(\theta)$ in the strip $-\pi/h<\im\theta<\pi/h$ (all related to
zeroes of the $Y_a(\theta)$ at the same points) we were able to recast the
generalised TBA equations, including the expressions for $c(r)$, into the
shape of the basic TBA equations, albeit with redefined functions
$\hat\ep_a$ and $\hat L_a$. This was very useful for the efficient
numerical treatment of the equations, and may be of deeper significance.
Note that the identities~(\ref{identi}) and (\ref{identt}) on which the
manoeuvre depends hold with trivial modifications
for all of the $ADET$-related diagonal scattering
theories, and so the technique is immediately applicable to a number of
other models.
\item 
{\bf Field theory and the generalised TBA: }
In all of the cases that we studied, it proved possible to extract exact
asymptotics from the generalised TBA equations which matched the
field-theoretic predictions found in refs.~\cite{La,KMd}.
These results relied on a number of intricate identities satisfied by the
mass spectra and S-matrices of the $T_N$-related models. An
outstanding problem is to set this into a broader context, and as a first
step it would be interesting to know the appropriate generalisations
to the other diagonal scattering theories. 
\item 
{\bf The crossover region: }
The detailed study made in sections~3 and~4 revealed that even for real
values of $r$ the generalised TBA equations can change in complicated ways
as one moves from the ultraviolet to the infrared, typically
undergoing a number of distinct transitions. These transitions can both
increase and decrease the number of active singularities, and as each is
passed, a new equation has to be solved. So far we can only appeal to a
case-by-case analysis, and more numerical work will probably be needed
before a general picture emerges. There is an intriguing
similarity between the
movements of singularities in the generalised TBA equations and the
movements of Bethe ansatz roots near to the scaling limit of lattice models,
as studied in ref.~\cite{GNa}; it remains to be seen whether any
significance should be attached to this observation.
\end{itemize} 

\noindent
One item of unfinished business is a study of more general states, even
in the $T_N$ models. Based on the experience with the SLYM discussed in
ref.~\cite{DTa}, it is not hard to guess how to include one-particle states
with non-zero spin: the key is to give up the $\theta\rightarrow-\theta$
symmetry of the pseudoenergies, and to replace each pair
$\{\theta^{(j)}_a,-\theta^{(j)}_a\}$ of active singularities in the
equations previously discussed with a pair
$\{\theta^{(j)}_a,\bar\theta^{(j)}_a\}$, imposing 
$\bar\theta^{(j)}_a=
(\theta^{(j)}_a)^*$ whenever $r$ is real. Note that these pairs of
singularities belong to either the left or right kink systems from the
outset, and so we expect there to be less need for them to undergo
transitions as the infrared region is left and the ultraviolet approached.
Preliminary studies of the
$T_2$ model show good agreement with results from truncated conformal
space, with, so far, no sign of any transitions at all,
through the whole range of
$r$. However the full story, both for these states and for states with
higher numbers of
particles, remains to be uncovered. Numerical problems are rather acute
when trying to track more than one or two active singularities, and indeed
it is becoming increasingly necessary to develop some
alternatives to the rather crude iteration schemes that we have been using
to date.

Finally we should mention a hope for the longer term, which is to develop an
overall understanding of the Riemann surfaces so far only
tentatively explored. There are many structures hidden in these
surfaces -- in particular we would emphasise 
their monodromy groups~\cite{Sb} -- and it is
natural to suppose that these will be related to
some of the other pieces of algebraic machinery that
enter into the study of integrable quantum field theories.

\medskip
\bigskip

\noindent{\bf Acknowledgments --- }We would like to thank  Sergei Lukyanov, 
John Parker and Francesco Ravanini  for useful discussions
at various stages, and everyone at
the Yukawa Institute, in particular Takeo Inami and Ryu Sasaki, for
their hospitality during the early stages of this work. In addition PED
would like to thank the Institute Henri Poincar\'e in Paris and the
Institute for Theoretical Physics in Santa Barbara for hospitality as the
project continued.
This work was supported in part by a Human
Capital and Mobility grant of the European Union, contract number
ERBCHRXCT920069, in part by a NATO grant, number CRG950751, in part by a
British Council/Royal Society/JSPS joint research project, and
in part by the National Science Foundation under
grant no.~PHY94-07194.
PED thanks the EPSRC for an Advanced Fellowship, and
RT thanks the Mathematics Department of Durham University for a
postdoctoral fellowship.

%
\appendix
\resection{Field-theoretic predictions for infrared asymptotics}
\label{ftres}
Finite-size corrections to the energies of one-particle states were
studied by L\"uscher in ref.~\cite{La}. To all orders in perturbation
theory, he found that the leading corrections could be
expressed in terms of universal quantities appearing in the
infinite-volume two-particle S-matrix elements. 
In this appendix we summarise the results of
\cite{La} as applied to the $T_N$ models, 
drawing also on the convenient and somewhat generalised
discussion provided by ref.~\cite{KMd}.

\begin{figure}[htb]
\epsfxsize=9cm
\epsfysize=7.5cm
{~~~~}\qquad\qquad\epsfbox{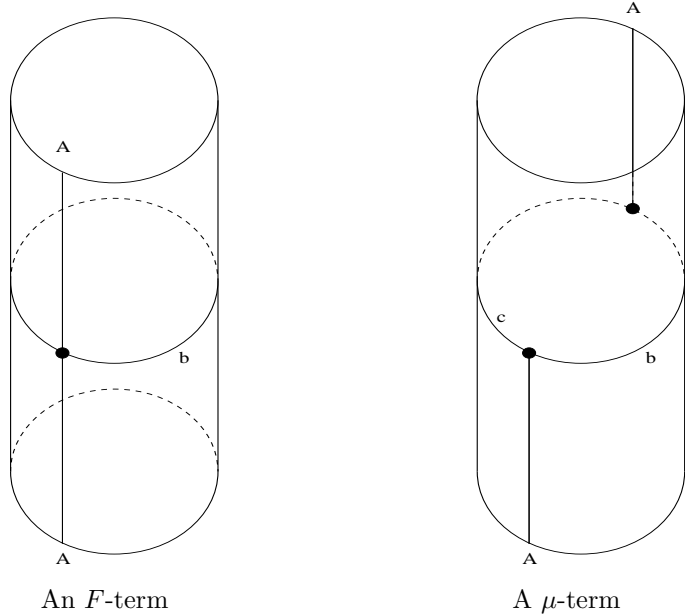}
\vskip 2pt
{\small\noindent\hfil 
An $F$-term\qquad\qquad\hfil\qquad A $\mu$-term\hfil\break}
\vskip -15pt
\caption{\leftskip=.5cm  {\protect \small The two classes of diagram
responsible for the leading contributions to $\Delta M_A\,$.}
\rightskip=0.5cm }  
\label{diags}
\end{figure}

If a zero-momentum one-particle state $A$ has a mass equal to $M_A$ 
on the infinite line, then the mass of the same state on a large but finite 
circle of circumference $R$ receives corrections from two main sources.
Those from the first, called $F$-terms in \cite{KMd}, give a total
contribution
\eq
\Delta_F M_A = -\sum_{b}{'~} {M_b\over 2\pi} {~\cal P} \iintd\,
e^{-M_bR\cosh\theta}(S_{Ab}(\theta{+}\imath\pi/2)-1)~.
\label{delta1}
\en
The prime indicates that terms smaller than the overall error can be
dropped (see ref.~\cite{KMd} for more discussion of this point)
and the ${\cal P}$ signals that the principal part
of the integral should be taken. (It is anyway irrelevant for the $T_N$
theories.) These terms can be traced to the interaction of the particle $A$ 
with a 
virtual particle $b$ travelling around the space-time cylinder, as
illustrated on the left of figure \ref{diags}.
\setlength{\unitlength}{1.mm}

Contributions of the second type (called $\mu$-terms in \cite{KMd}) 
are due to the processes illustrated on the right of figure~\ref{diags},
in which $A$ splits into a pair of virtual 
particles $b$ and $c$, which travel around the world before recombining
to form  $A$ again. The result is proportional to the squares of the
on-shell three-point couplings $f_{Abc}$, and these can be determined
from the residues $\R S_{Ab}(\imath U^c_{Ab})$ of the
two-particle S-matrix elements at the appropriate
fusing angles $\imath U^c_{Ab}$. 
The total shift induced by processes of this sort is
\eq
\Delta_{\mu} M_{A} = -\sum_{b,c}{}'~\Theta(M_{A}^2-|M_b^2-M_c^2|)\,
\mu_{Abc} R_{Abc} e^{-\mu_{Abc}R} 
\label{delta2}
\en
where
\eq
R_{Abc}= -\imath M_{Abc}\R S_{Ab}(\imath U^{c}_{Ab})\,,
\en
and 
\eq
\mu_{Abc}= M_b\sin U^{c}_{Ab}=M_c\sin U^{b}_{Ac}=
{M_bM_c\over M_A}\sin U^{A}_{bc}~.
\en
In these definitions,
$M_{Abc}$ is one if $f_{Abc}\neq 0$ and zero otherwise, and the step
function
\[
\Theta(x)=  \left\{ \begin{array}{l}
  0          \mbox{~~if~~}  x<0 \\
 {1 \over 2} \mbox{~~if~~}  x=0 \\
  1          \mbox{~~if~~}  x>0 \\
\end{array} \right.
\]
serves to disqualify those virtual processes $A\rightarrow b\,c\rightarrow A$
which cannot be drawn as on-shell diagrams on the cylinder. 

\setlength{\unitlength}{1.mm}
\begin{figure}[tbh]
\epsfxsize=11cm
\epsfysize=6.5cm
{~~~~}\epsfbox{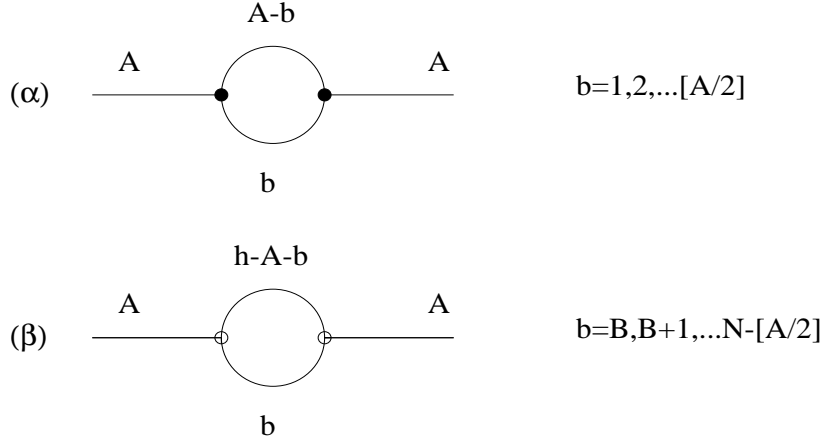}
\caption{\leftskip=.5cm  {\protect \small  
A set of $\mu$-terms in the $T_N$ theory.}
\rightskip=0.5cm }  
\label{loop}
\end{figure}
Specialising to the non-unitary $T_N$ theories, the non-vanishing
couplings are as follows:
\[
\begin{array}{l}
\openup 1\jot
f_{Abc}\in~\RR^* \mbox{~~if~~}c=A{+}b \le N\mbox{~or~}c=|A{-}b|\neq 0\,;\\
f_{Abc}\in\imath\RR^* \mbox{~~if~~} c=h-(A{+}b)\le N\,;\\
f_{Abc}=~0 \mbox{~~~~otherwise,~~}
\end{array}
\]
and the fusing angles are
\bea
\openup 1\jot
U^{|A-b|}_{Ab}&=&\lf(1 - {|A-b| \over h}\ri)\pi\,;\nn\\
U^{n(A,b)}_{Ab}&=&\lf({A+b \over h}\ri)\pi\,,\nn
\eea
with $n(A,b)=\min(A{+}b,h{-}A{-}b)$. 
Due to the step function in (\ref{delta2}), only the  
contributions corresponding to the diagrams represented in 
figure~\ref{loop} are non-vanishing, and these give 
\bea
\mu_{A,b,A-b}&=& m_b \sin(\fract{|A{-}b|}{h}\pi)\,,\nn\\
R_{A,b,A-b}&=& -\imath\R S_{Ab}(\imath(1-\fract{|A{-}b|}{h})\pi),
\quad b=1,2,\dots [A/2]\quad
\label{muone}
\eea
for $\alpha$-processes, and
\bea
\mu_{A,b,h-A-b}&=&m_b \sin(\fract{(A{+}b)}{h}\pi)\,,\nn\\
R_{A,b,h-A-b}&=& -\imath\R S_{Ab}(\imath\fract{|A{+}b|}{h}\pi),
\quad b=B,B{+}1,\dots N{-}[A/2] 
\label{mutwo}
\eea
for $\beta$-processes.
Note that the two types of process contribute oppositely to
$\Delta_{\mu}M_A$, since the couplings involved in the $\alpha$-processes
are real, while those in the $\beta$-processes are purely imaginary.
The total number of distinct contributions is equal to $A$, which is
reflected in
the fact that there are $A$ independent active singularities in the
relevant generalised TBA equation.

\bigskip\bigskip\bigskip

\resection{Numerical results}
\label{numapp}
In this appendix, numerical solutions to the excited-state TBA equations
discussed in section 3 are compared with TCSA data. 
For the first one-particle state,
table~\ref{oneTBAuv} was compiled using equations (\ref{oneTBAg}) --
(\ref{oneTBAgt}), and table~\ref{oneTBAir} using equations 
(\ref{oneTBA}), (\ref{oneTBAc}) and (\ref{oneTBAt}). For the second
one-particle state, table~\ref{tabtwouv} used using
(\ref{modaa}), (\ref{modbb}), and tables \ref{tabtwomed} 
and~\ref{tabtwoir}, (\ref{twomedTBAdefa}) -- (\ref{twoTBAgt}).
Table~\ref{tabtwoir} was also checked using (\ref{R11}) -- (\ref{t2q}).
Accuracies for the TBA results were estimated by varying parameters such
as the discretisation of the $\theta$ axis until the results stabilised
at the quoted values, whilst accuracies for the TCSA were estimated by
varying the
truncation level. 
The first TBA entry in table~\ref{tabtwomed} was obtained by extrapolation
from larger values of $r$,
since our iterative routine failed to converge there.
\begin{table}[bh]
\widetable
\begin{center}
\begin{tabular}[t]{|c|l|l|l|}  
\hline \hline
\rll
{}~r~{} &~~~~ TBA~~~ &~~~~~ TCSA~~~~~~    &~~~~$\beta_2^{(0)}$ ~~~~ \\
\hline
 $0.25$ & $ 0.23816052575(2)$& $ 0.23816052577(4)  $ & $2.44086172191(0)$ \\
 $0.50$ & $ 0.23856619676(8)$& $ 0.2385661967(3)   $ & $1.74875349244(1)$ \\
 $0.75$ & $ 0.23957858011(5)$& $ 0.23957857(9)     $ & $1.34585654321(1)$ \\
 $1.00$ & $ 0.24140149459(4)$& $ 0.24140149(2)     $ & $1.06270397320(0)$ \\
 $1.25$ & $ 0.24415555107(8)$& $ 0.2441555(4)      $ & $0.84614229277(0)$ \\
 $1.50$ & $ 0.24785826484(5)$& $ 0.2478582(4)      $ & $0.67206426675(5)$ \\
 $1.75$ & $ 0.25241760010(3)$& $ 0.252417(5)       $ & $0.52676984074(1)$ \\
 $2.00$ & $ 0.25764227087(8)$& $ 0.257642(1)       $ & $0.40064138633(5)$ \\
 $2.25$ & $ 0.26326448567(0)$& $ 0.263264(2)       $ & $0.28454565858(0)$ \\
 $2.50$ & $ 0.26896728571(2)$& $ 0.268967(1)       $ & $0.16239328864(9)$ \\
\hline \hline
\end{tabular}
\end{center}
\caption{\leftskip=.5cm  {\protect \small First one-particle
level, $r<r_c\,.$} \rightskip=0.5cm }  
\label{oneTBAuv}
\end{table}
%
\begin{table}[h]
\widetable
\begin{center}
\begin{tabular}[t]{|c|l|l|l|}  
\hline \hline
\rll
{}~r~{} &~~~~ TBA~~~  &~~~ TCSA~~   &~~~$\imath\,\beta_2^{(0)}$ ~~~~ \\
\hline
 $2.75$ & $ 0.27440978264(3) $& $ 0.274409(3) $ & $0.10660570253(6)$ \\
 $3.00$ & $ 0.27924699310(4) $& $ 0.279246(3) $ & $0.19389451747(5)$ \\
 $3.25$ & $ 0.28314394898(6) $& $ 0.283143(1) $ & $0.23644517085(6)$ \\
 $3.50$ & $ 0.28578526162(5) $& $ 0.28578(4)  $ & $0.26216147499(3)$ \\
 $3.75$ & $ 0.28688156615(0) $& $ 0.28688(0)  $ & $0.27879066486(6)$ \\
 $4.00$ & $ 0.28617389043(3) $& $ 0.28617(2)  $ & $0.28989626682(7)$ \\
 $4.25$ & $ 0.28343648042(0) $& $ 0.28343(4)  $ & $0.29744353109(2)$ \\
 $4.50$ & $ 0.27847823059(8) $& $ 0.27847(6)  $ & $0.30262252477(1)$ \\
 $4.75$ & $ 0.27114269310(0) $& $ 0.27114(0)  $ & $0.30619467150(4)$ \\
 $5.00$ & $ 0.26130666042(9) $& $ 0.26130(4)  $ & $0.30866417603(4)$ \\
 $5.25$ & $ 0.24887745081(7) $& $ 0.24887(5)  $ & $0.31037226542(4)$ \\
 $5.50$ & $ 0.23378918672(7) $& $ 0.23378(7)  $ & $0.31155300241(8)$ \\
 $5.75$ & $ 0.21599846946(4) $& $ 0.21599(6)  $ & $0.31236819753(5)$ \\
 $6.00$ & $ 0.19547988038(6) $& $ 0.19547(9)  $ & $0.31293014812(8)$ \\
 $7.00$ & $ 0.08602148615(6) $& $ 0.08602(7)  $ & $0.31388982426(3)$ \\
 $8.00$ & $-0.06688310507(3) $& $-0.0668(6)   $ & $0.31410090210(8)$ \\
 $10.0$ & $-0.50050238863(1) $& $-0.500(4)    $ & $0.31415656313(2)$ \\
 $15.0$ & $-2.31926770945(7) $& $-2.31(8)     $ & $0.31415926412(7)$ \\
 $20.0$ & $-5.18417167922(3) $& $-5.18(2)     $ & $0.31415926535(8)$ \\
 $30.0$ & $-14.0517103842(1) $& $-14.0(4)     $ & $0.31415926535(8)$ \\
\hline \hline
\end{tabular}
\end{center}
\caption{\leftskip=.5cm  {\protect \small First one-particle level, 
$r>r_c\,.$} \rightskip=0.5cm }  
\label{oneTBAir}
\end{table}
\begin{table}[h]
\widetable
\begin{center}
\begin{tabular}[t]{|c|l|l|l|}  
\hline \hline
\rll
{}~r~{} &~~~~ TBA~~~&~~~~~ TCSA~~~~~~ &~~~~$\beta_1^{(0)}$ ~~~~ \\
\hline
$0.25$ & $0.8095238098(2) $& $0.80952380986(1)$ & $3.2949747170(7)$  \\   
$0.50$ & $0.8095238271(8) $& $0.8095238272(2) $ & $2.6018275552(3)$  \\    
$0.75$ & $0.8095239881(9) $& $0.8095239882(1) $ & $2.1963626206(8)$  \\   
$1.00$ & $0.8095247257(6) $& $0.8095247258(0) $ & $1.9086813430(7)$  \\   
$1.25$ & $0.8095270410(0) $& $0.8095270410(7) $ & $1.6855402842(2)$  \\  
$1.50$ & $0.8095327731(5) $& $0.809532773(2)  $ & $1.5032248871(9)$  \\   
$1.75$ & $0.8095448001(5) $& $0.809544800(3)  $ & $1.3490870895(8)$  \\   
$2.00$ & $0.8095670652(3) $& $0.809567065(4)  $ & $1.2155794073(0)$  \\   
$2.25$ & $0.8096043028(5) $& $0.80960430(4)   $ & $1.0978356190(0)$  \\   
$2.50$ & $0.8096613348(7) $& $0.80966133(5)   $ & $0.9925341024(8)$  \\   
$2.75$ & $0.8097417885(7) $& $0.80974178(9)   $ & $0.8973043325(8)$  \\   
$3.00$ & $0.8098461844(7) $& $0.80984618(3)   $ & $0.8103904953(7)$  \\   
$3.25$ & $0.8099693378(1) $& $0.80996933(8)   $ & $0.7304470394(3)$  \\   
$3.50$ & $0.8100972214(6) $& $0.81009722(2)   $ & $0.6564058765(0)$  \\  
$3.75$ & $0.8102034874(6) $& $0.8102034(9)    $ & $0.5873835438(9)$  \\
$4.00$ & $0.810246050(1)  $& $0.8102460(6)    $ & $0.522609957(4)$  \\
$4.25$ & $0.810164185(2)  $& $0.810164(2)     $ & $0.461365983(5)$  \\
$4.50$ & $0.809876615(7)  $& $0.809876(7)     $ & $0.402917258(4)$  \\ 
\hline \hline 
\end{tabular}
\end{center}
\caption{\leftskip=.5cm  {\protect \small Second one-particle
level, $r<r_{c_1}\,.$} \rightskip=0.5cm }
\label{tabtwouv}
\end{table}
%
\begin{table}[h]
\widetable
\begin{center}
\begin{tabular}[t]{|c|l|l|l|l|}  
\hline \hline
\rll
{}~r~{} &~~ TBA~~ & TCSA     &~~~$\imath\,\beta_1^{(0)}$ ~~&~~
$-\imath\,\theta_2^{(0)}$  \\ 
\hline                   
$4.75$ & $0.8092809(5) $& $0.80928(1)  $ & $0.346424(5)$ &
$0.015804(6)$  \\
$5.00$   & $0.80825467(3)   $& $0.80825(5)  $ & $0.29078538(8)$   &
0.10280614(0)  \\
$5.25$   & $0.80665753(1)   $& $0.80665(8)  $ & $0.23427070(8)$   &
0.14891716(3)  \\
$5.50$   & $0.80433518(4)   $& $0.80433(6)  $ & $0.17338533(0)$   &
0.18056955(6)  \\
$5.75$   & $0.80112353(4)   $& $0.80112(4)  $ & $0.09649407(3)$   &
0.20409098(1)  \\ 
\hline \hline 
\end{tabular}
\end{center}
\caption{\leftskip=.5cm  {\protect \small Second one-particle
level, $r_{c_1}<r<r_{c_2}\,.$} \rightskip=0.5cm }  
\label{tabtwomed}
\end{table}
%
\begin{table}[h]
\widetable
\begin{center}
\begin{tabular}[t]{|c|l|l|l|l|}  
\hline \hline
\rll
{}~r~{} &~~ TBA~~ & TCSA     &~~~$\imath\,\beta_1^{(0)}$ ~~&~~
$-\imath\,\theta_2^{(0)}$  \\ \hline                   
$6.00$ & $ 0.7968556702(0) $& $0.79685(7)$ & $0.0896301999(2)$ & 
$0.2223149258(0)$  \\
$6.25$ & $ 0.7913547600(8) $& $0.79135(9) $ & $0.1501055922(9)$ & 
$0.2368163134(7)$  \\
$6.50$ & $ 0.7844612122(0) $& $0.78446(8) $ & $0.1861508713(2)$ & 
$0.2485780614(9)$  \\
$6.75$ & $ 0.7760128576(7) $& $0.7760(2) $ & $0.2116358291(4)$ & 
$0.2582559981(1)$  \\
$7.00$ & $ 0.7658591165(7) $& $0.7658(7) $ & $0.2308123309(7)$ & 
$0.2663096804(3)$  \\
$7.25$ & $ 0.7538605345(2) $& $0.7538(8) $ & $0.2457260952(8)$ & 
$0.2730722975(5)$  \\
$7.50$ & $ 0.7398899544(4) $& $0.739(9)  $ & $0.2575655998(7)$ & 
$0.2787920410(3)$  \\
$7.75$ & $ 0.7238331001(3) $& $0.723(8)  $ & $0.2670979493(9)$ & 
$0.2836579633(7)$  \\ 
$8.00$ & $ 0.7055886730(5) $& $0.705(6)  $ & $0.2748522116(1)$ & 
$0.2878168593(0)$  \\
$8.50$ & $ 0.6621947642(6) $& $0.662(2)  $ & $0.2864555943(6)$ & 
$0.2944542899(1)$  \\
$9.00$ & $ 0.6091397136(7) $& $0.609(2)  $ & $0.2944284302(9)$ & 
$0.2993877628(7)$\\
$10.0$ & $ 0.4725672915(3) $& $0.472(7)  $ & $0.3039279521(5)$ & 
$0.3058454231(1)$\\
$15.0$ & $-0.8450264306(8) $& $-0.84(3)   $ & $0.3136882646(6)$ & 
$0.3137048805(9)$  \\
$20.0$ & $-3.2169979044(2) $& $-3.21(5)   $ & $0.3141349870(6)$ & 
$0.3141351300(9)$  \\
$30.0$ & $-11.100815854(4) $& $-1(4)      $ & $0.3141591977(4)$ & 
$0.3141591977(4)$  \\ \hline \hline
\end{tabular}
\end{center}
\caption{\leftskip=.5cm  {\protect \small Second one-particle
level, $r>r_{c_2}\,.$} \rightskip=0.5cm }  
\label{tabtwoir}
\end{table}
%

\end{document}